\documentclass[preprint]{imsart}

\RequirePackage{amsthm,amsmath}
\usepackage[round]{natbib}
\RequirePackage[colorlinks,citecolor=blue,urlcolor=blue]{hyperref}

\doi{10.1214/154957804100000000}
\pubyear{0000}
\volume{0}
\firstpage{1}
\lastpage{1}

\startlocaldefs

\usepackage{amssymb}
\usepackage{amsfonts}
\usepackage{multirow}
\usepackage{todonotes}
\usepackage{mathrsfs} 
\usepackage{algorithm}
\usepackage{algorithmic}
\usepackage{verbatim}

\newtheorem{theorem}{Theorem}
\newtheorem{lemma}{Lemma}
\newtheorem{corollary}{Corollary}
\newtheorem{proposition}{Proposition}
\newtheorem{definition}{Definition}
\theoremstyle{remark}

\newtheorem{remark}{Remark}

\newcommand{\ep}{\epsilon}
\newcommand{\de}{\delta}
\newcommand{\F}{\hat F}
\newcommand{\iid}{\overset {\text{i.i.d}}{\sim}}
\newcommand{\mscr}[1]{\mathscr #1}
\newcommand{\RR}{\mathbb{R}}
\newcommand{\red}{}

\newcommand{\HH}{\mathrm{H}}

\newcommand{\twid}{\widetilde}

\endlocaldefs

\begin{document}
\begin{frontmatter}

\title{Differentially Private Kolmogorov-Smirnov-Type Tests\support{This work was supported in part by NSF Award No. SES-2150615 to Purdue University.}}
\runtitle{DP KS-Type Tests}


\author{\fnms{Jordan} \snm{Awan}\corref{}\ead[label=e1]{jawan@purdue.edu}}
\address{Department of Statistics\\Purdue University\\West Lafayette, IN 47907\\\printead{e1}}
\and
\author{\fnms{Yue} \snm{Wang}\ead[label=e2]{yw3930@columbia.edu}}
\address{Department of Statistics\\ Columbia University\\ New York, NY 10027\\\printead{e2}}

\runauthor{J. Awan and Y. Wang}

\begin{abstract}
Hypothesis testing is a central problem in statistical analysis, and there is currently a lack of differentially private tests which are both statistically valid and powerful. In this paper, we develop several new differentially private (DP) nonparametric hypothesis tests. Our tests are based on 
Kolmogorov-Smirnov, Kuiper, Cram\'er-von Mises, and Wasserstein test statistics, which can all be expressed as a pseudo-metric on empirical cumulative distribution functions (ecdfs), and can be used to test hypotheses on goodness-of-fit, two samples, and paired data. 
We show that these test statistics have low sensitivity, requiring minimal noise to satisfy DP. In particular, we show that the sensitivity of these test statistics can be expressed in terms of the \emph{base sensitivity}, which is the pseudo-metric distance between the ecdfs of adjacent databases and is easily calculated. The sampling distribution of our test statistics are distribution-free under the null hypothesis, enabling easy computation of $p$-values by Monte Carlo methods. 
We show that in several settings, especially with small privacy budgets or heavy-tailed data, our new DP tests outperform alternative nonparametric DP tests.
\end{abstract}





\end{frontmatter}


\section{Introduction}
Developing powerful nonparametric/distribution-free hypothesis tests is important as data do not always come from known distributions. Such tests may be employed by themselves, or as one step in a broader statistical analysis. As data privacy concerns arise, such as in medical settings, social science research, or big tech companies, there is a growing need to develop nonparametric tests that also offer a formal privacy guarantee.

Differential privacy (DP) introduced in \citet{dwork2006calibrating}, provides rigorous privacy protections for database participants. DP methods require the introduction of additional randomess in the analysis procedure, which obscures the contribution of one individual in the dataset. The randomness is designed in a way that the distribution of possible outputs from the DP algorithm is similar if one person's data is changed in the dataset, offering a notion of plausible deniability. The level of privacy is characterized by the \emph{privacy budget} $\epsilon$, where smaller values of $\ep$ ensure stronger privacy guarantees. DP methods are being widely employed by companies such as Google \citep{erlingsson2014rappor}, Apple \citep{tang2017privacy}, and Microsoft \citep{ding2017collecting}, as well as by the US Census \citep{abowd2018us}.

There are two main challenges when developing DP hypothesis tests: 1) the noise required for DP must be scaled according to the \emph{sensitivity} of the test statistic. Many test statistics have either very high or even unbounded sensitivity {\red (e.g., t-test \citep{couch2019}, F-statistic \citep{alabi2022hypothesis}, Anderson-Darling (\ref{rem:anderson}))}, requiring either a large amount of noise to satisfy DP, or even requiring a modification to the test statistic to reduce the sensitivity and 2) the sampling distribution of the noisy test statistic must be derived to ensure accurate type I errors and $p$-values. While the noise for DP is often asymptotically negligible, it has been demonstrated that traditional asymptotic methods {\red based on convergence in distribution} have unacceptable accuracy in realistic sample sizes \citep{wang2018statistical}. 

To address both challenges, we identify a family of test statistics, based on empirical cumulative distribution functions (empirical cdfs/ecdfs) which have low sensitivity, and  whose sampling distributions are distribution-free under the null hypothesis. Due to the low sensitivity, these test statistics can be privatized using a small amount of noise; due to the distribution-free nature of these test statistics, we can derive accurate $p$-values by either Monte Carlo simulations or by using the asymptotic approximations of \citet{wang2018statistical}, which are guaranteed to be at least as accurate as the non-private approximations. {\red To facilitate the calculation of sensitivity for these test statistics, we propose using an intermediate calculation called \emph{base sensitivity}, that captures the distance between the ecdfs of adjacent databases and which is usually easily calculated. Through simulations we show that our proposed tests outperform existing DP tests in several settings, especially with either small privacy budgets or heavy-tailed data, and that they have comparable performance to prior methods in other settings.} \\

\noindent{\bf Organization: }In Section \ref{s:background}, we set the notation for the paper, and review the necessary background on hypothesis testing and differential privacy. In particular, we discuss how the noise for DP must be scaled according to the sensitivity of a statistic. In Section \ref{s:sensitivity} we describe the family of test statistics that we consider, which can all be expressed as a metric (or pseudo-metric) on the space of cdfs. In particular, we consider the Kolmogorov-Smirnov (KS), Kuiper, Cram\'er-von Mises, and Wasserstein tests. In Section \ref{s:sensitivity}, we introduce a concept called \emph{base sensitivity}, which measures the distance between two ecdfs for adjacent databases, in terms of the metric in the test statistic; for all of the pseudo-metrics considered in this paper, we show that the base sensitivity is $1/n$. In Sections \ref{s:gof}-\ref{s:paired}, we show that for goodness-of-fit, two-sample, and symmetry/paired data applications, the sensitivity of the test statistic can be expressed in terms of the base sensitivity, and that the test statistics require minimal noise to achieve DP. All of our DP tests are distribution-free, and accurate $p$-values can easily be computed using Monte Carlo methods.  In particular, in Section \ref{s:gof}, we show that the KS and Kuiper goodness-of-fit tests are distribution-free for models with unknown location and scale parameters, a result that may be of interest to the broader statistical community. In Sections \ref{s:goodSim}, \ref{s:2simulations}, and \ref{s:pairedSimulation} we compare our DP tests against each other, as well as against other competing tests in the DP literature, through simulations. All proofs are deferred to Appendix \ref{app:proofs}. 

For simplicity of presentation, we state our results in terms of $\epsilon$-DP. However, as all of our private tests simply add noise to a non-private test statistic, the tests are easily adapted to other forms of DP (e.g., approximate-DP, concentrated-DP, Gaussian-DP, etc.) by adding a different form of noise. This is discussed in more detail in Section \ref{s:discussion}.\\

\noindent{\bf Related work: }\citet{couch2019} proposed several non-parametric hypothesis tests that satisfy differential privacy and compare the performance against each other, including  privatized versions of  Wilcoxon signed rank, Mann-Whitney, and Kruskal-Wallis tests. {\red Of these tests, their privatized  Wilcovon test is a significant improvement over a previous DP Wilcoxon test developed by \citet{task2016differentially}, while the others are the first DP implementations of these tests.} \citet{awan2018differentially} derived \emph{uniformly most powerful} (UMP) test for binary data among all DP $\alpha$-level tests, as well as DP versions of the sign test and median test. 
In the follow-up paper, \citet{awan2020differentially} extended these results to allow for optimal two-sided tests as well as the construction of optimal confidence intervals. \citet{gaboardi2016differentially} developed private goodness-of-fit and independence tests for categorical data, based on $\chi^2$-statistics. \citet{awan2023canonical} developed differentially private tests for the difference of population proportions. In parametric models, \citet{ferrando2022parametric} propose using the parametric bootstrap to produce differentially private confidence intervals; this approach could also be used to perform hypothesis tests on parametric families. {\red \citet{awan2023simulation} applied simulation-based inference techniques to produce confidence sets and hypothesis tests on privatized data with guaranteed coverage/type I error. } \citet{wang2018statistical} developed an asymptotic framework to derive statistical approximating distributions under differential privacy, which they show can be used to develop differentially private $\chi^2$ goodness-of-fit tests, tests for independence of multinomials, and goodness-of-fit via the Kolmogorov-Smirnov test. \citet{yu2018differentially} developed a paired-data DP Kolmorov-Smirnov test for the purpose of comparing predicted values from linear regression models to the true data points.

{\red \citet{drechsler2022nonparametric} developed private confidence intervals for the median, and used a privatized empirical cdf as part of these mechanisms; these intervals could be inverted to produce private nonparametric hypothesis tests for the median. Some other works that produce privatized empirical cdfs include \citet{hay2010boosting} and \citet{honaker2015efficient}. \citet{bi2023distribution} also produce private distribution estimators, however their method requires the use of a ``hold-out dataset,'' which is not formally protect, and so the resulting method is not fully differentially private. 

Among the above related work, the most relevant competitors to our proposed methods are the sign test and median test \citep{awan2018differentially} and the Kruskal Wallis, Mann Whitney, and Wilcoxon tests \citep{couch2019}, as these are the nonparametric tests which can be applied in the same settings as ours. Similar to the tests developed in this paper, all of these DP tests rely on sensitivity calculations and noise-addition to achieve DP. Throughout the paper, we compare the performance of our proposed tests against these competitors through simulations to determine their relative efficiency in various scenarios.}


\section{Notation and background}\label{s:background}
In this section, we set the notation for the paper and review background on hypothesis testing and differential privacy. 

Let $F$ be a cumulative distribution function (cdf). We write $x\sim F$ to denote that $x$ is a random variable with distribution $F$. For a vector $(x_1,\ldots, x_n)$, we write $x_i\iid F$ to denote that each $x_i$ is independently and identically distributed (i.i.d.) with distribution $F$. 

For a sample $(x_i)_{i=1}^n$, the empirical cdf (ecdf) is $\F_x(t)=n^{-1} \sum_{i=1}^n I(x_i\leq t)$. We assume throughout the paper that the true cdf $F$ is continuous, which will be important to ensure that our proposed tests are distribution-free.

Given a space $\mscr X$, a \emph{pseudo-metric} is a function $d:\mscr X\times \mscr X\rightarrow \RR^{\geq 0}$, which satisfies 1) $d(x,x)=0$, 2) $d(x,y)=d(y,x)$, and 3) $d(x,y)\leq d(x,z)+d(z,y)$ for all $x,y,z\in \mscr X$. A \emph{metric} is a pseudo-metric with the additional property that $d(x,y)=0$ implies that $x=y$.

\subsection{Hypothesis testing}
Let $x=(x_1,\ldots, x_n)\in \mscr X^n$ be distributed $x_i \stackrel{iid}{\sim} F_\theta$, where $\theta\in \Theta$. Let $\Theta_0, \Theta_1$ be disjoint subsets of $\Theta$. We call  $\Theta_0$ the \emph {null} and $\Theta_1$ the \emph{alternative}. A \emph{(randomized) test} of $H_0: \theta \in \Theta_0$ versus $H_1: \theta\in \Theta_1$ is a measurable function $\phi: \mscr X^n \rightarrow [0,1]$. We say a test $\phi$ is at \emph{level} $\alpha$ if $\sup_{\theta\in \Theta_0} E_{ F_\theta} \phi \leq \alpha$. The choice of $\alpha$ determines the \emph{type I error}, which is the probability of rejecting the \emph{null} hypothesis when it is correct. We define the \emph{power} as the probability of rejecting the null hypothesis when it is false.

A $p$-value is a (randomized) function $p(x)$ taking values in $[0,1]$, which satisfies the condition $\sup_{\theta\in \Theta_0}P(p(x)\leq t)\leq t$. A $p$-value can be interpreted as the smallest type I error at which the null hypothesis could be rejected, given the data $x$. 

Suppose that $T(x)$ is a (randomized) test statistic such that larger values give more evidence for $\Theta_1$. If the distribution of $T(x)$ is the same for all $\theta\in \Theta_0$, we say that $T$ is \emph{distribution-free}, and a $p$-value is $1-F_T(T)$, where $F_T$ represents the distribution of $T(x)$ under $\Theta_0$. For distribution-free tests, one may 1) analytically derive the distribution $F_T$, 2) approximate $F_T$ under the null hypothesis by Monte Carlo, or 3) use asymptotic techniques to approximate the distribution $F_T$. In this paper, we will use Monte Carlo methods as deriving the exact form is intractable, and Monte Carlo allows us to incorporate a possible dependence on the sample size, that asymptotic methods do not.


\subsection{Differential privacy}

 \emph{Differential Privacy} (DP) is a framework that enables researchers to publish population-level information from a database, without divulging individuals' personal information. For any two datasets that differ in one person's data, DP requires that the probability of producing any set of outputs is within a factor of $\exp(\epsilon)$. The privacy budget $\epsilon$ controls how much the output of the mechanism can differ between the two adjacent databases. For a small value of $\epsilon$, the two mechanisms become more and more similar offering more privacy. A large value of $\epsilon$ will lead to more accurate computations but weak privacy protection.
 
 Differential privacy requires a metric to quantify when two databases are differing in one entry. Let $\mscr X$ be the space of possible inputs from one person, and let $\mscr X^n$ denote the set of possible databases with $n$ entries. We call a metric $m$ on $\mscr X^n$ an \emph{adjacency metric}, if for $x,x'\in \mscr X^n$, $m(x,x')\leq 1$ represents that $x$ and $x'$ differ in one entry. In this case, we call $x$ and $x'$ \emph{adjacent}. A common choice for $m$ is the Hamming distance, which counts how many entries $x$ and $x'$ differ in: $\HH(x, x' ) = \# \{x_i\neq x'_i\}$. 

\begin{definition}
[Differential Privacy]
A \emph{mechanism} $\mathscr{P}_x$ is a set of distributions indexed by the space of possible databases $\mscr X^n$. For a given adjacency metric $m$ on $\mscr X^n$, we say the mechanism $\mathscr{P}_x = \{P_x  \mid  x \in \mscr{X}^n\}$ satisfies $\epsilon$-differential privacy ($\epsilon$-DP) if for all sets of outputs $B$ and all $x, x' \in \mscr X^n$ such that $m(x,x')\leq 1$, we have $P_{x}(B) \leq \exp(\epsilon) P_{x'}(B)$ 
\end{definition}

Differentially private mechanisms are immune to \emph{post-processing} \citep[Proposition 2.1]{dwork2014algorithmic}, meaning that applying any data-independent procedure to an $\ep$-DP output does not change the $\ep$-DP guarantee. For the purposes of this paper, if a test statistic $T$ satisfies $\epsilon$-DP then the calculation of $p$-values from $T$, in a data-independent manner, also satisfies $\epsilon$-DP. 

 The \emph{sensitivity} of a function measures how much the function value changes when one person's data in the dataset is changed, capturing the influence that a single individual can have on the statistic. Sensitivity is a crucial concept in differential privacy, as the amount of noise added to achieve DP is scaled proportional to the sensitivity of the statistic. 

\begin{definition}
[Sensitivity]\label{def:sensitivity}
The sensitivity for a function $f: \mathscr{X}^n \rightarrow \mathbb{R}$, with respect to the adjacency metric $m$ is:
\[\Delta = \sup_{m(x,x')\leq 1} |f(x)-f(x')|,\] 
where the supremum is over any two adjacent databases.
\end{definition}

 A simple way to achieve DP is to add noise scaled proportional to the sensitivity. A very common method of doing so is the Laplace mechanism \citep{dwork2006calibrating}. Note that the same adjacency metric must be used in the sensitivity calculation as in the DP definition.

\begin{proposition}
[Laplace Mechanism: \citealp{dwork2006calibrating}]\label{prop:laplace}
Let $f: \mscr X^n \rightarrow \mathbb{R}$ be any function, the Laplace Mechanism is defined as:
\begin{equation}
\begin{split}
\Tilde{f}(x) = f(x) + L,
\end{split}
\end{equation}
where $L\sim\mathrm{Laplace}(\Delta/\epsilon)$ and $\Delta$ is the global sensitivity of $f$. Then, $\Tilde{f}$ satisfies  $\epsilon$-DP. Recall that the density of Laplace is $\mathrm{Lap}(x|b)= (2b)^{-1}\exp(-|x|/b)$.
\end{proposition}

 \citet{awan2018differentially} proposed the \emph{Truncated-Uniform-Laplace (Tulap)} mechanism, which in the case of $\epsilon$-DP (as opposed to $(\epsilon,\delta)$-DP) is an instance of the \emph{staircase mechanism} \citep{geng2015optimal}. {\red The Tulap distribution is generated by the convolution of a discrete Laplace random variable with a uniform random variable, and the combination is then truncated within a region centered around zero. The Tulap mechanism can be used to satisfy either $\ep$-DP or the more general $(\ep,\de)$-DP, depending on the truncation parameter.}  
 \citet{awan2023canonical} showed that Tulap is a \emph{canonical noise distribution}, meaning that it is precisely tailored to the privacy guarantee $\epsilon$-DP. 

\begin{proposition}
[Tulap Mechanism: \citealp{awan2018differentially}]\label{prop:tulap}
Let $f:\mscr X^n\rightarrow \RR$. The Truncated-Uniform-Laplace(Tulap) Mechanism is defined as:
\begin{equation}
\begin{split}
\Tilde{f}(x) = f(x) + \Delta  T,
\end{split}
\end{equation}
where $T\sim\mathrm{Tulap}(\exp(-\epsilon),0)$ and $\Delta$ is the global sensitivity of f. Then $\Tilde{f}$ satisfies $\epsilon$-DP. 
\end{proposition}

{\red \citet{awan2018differentially} showed that a Tulap random variable $\mathrm{Tulap}(b,0)$ can be easily generated as follows:} Let $U\sim U(-1/2,1/2)$, $G_1,G_2\iid \mathrm{Geom}(1-b)$, where $P(G_1=x)=b(1-b)^x$ for $x=0,1,2,\ldots$. Then $T\overset d=U+G_1-G_2\sim \mathrm{Tulap}(b,0)$. 
The R package \texttt{binomialDP} includes a sampler for the general Tulap distribution \citep{binomialDP}.

\begin{remark}
As Laplace and Tulap are two viable options of adding noise to achieve DP, one may wonder if one always outperforms the other. Through preliminary simulations, we found that for a statistic with sensitivity 1, Tulap performed better when the statistic is integer-valued, and Laplace performed better when this is not the case. The intuition behind this is that when the statistic is integer-valued, the Tulap mechanism is equivalent to the geometric/discrete Laplace mechanism, which has certain optimality properties for count statistics \citep{ghosh2009universally}.
\end{remark}

\section{Sensitivity of empirical CDF test statistics}\label{s:sensitivity}

Many test statistics are based on the ecdf, such as Kolmogorov-Smirnov, Cram\'er-von Mises, Wasserstein, and Kuiper tests \citep{stephens1974edf}. In fact, these tests are often framed in terms of a pseudo-metric $d$ applied to the ecdf, and can be used to test goodness-of-fit, two samples, paired data, and symmetry. In this paper, we are interested in computing the sensitivity (as defined in Defintion \ref{def:sensitivity}) of these test statistics. In order to simplify these sensitivity calculations, we propose a new concept, \emph{base sensitivity}, which only measures the distance, in terms of the relevant pseudo-metric, between two adjacent ecdfs. In the following subsections, we show that for all of the pseudo-metrics considered in this paper, the base sensitivity is $1/n$. In Sections \ref{s:gof}-\ref{s:paired}, we will see that the sensitivity of the test statistics can be expressed in terms of the base sensitivity, requiring a minimal amount of noise to achieve DP. 

For a pseudo-metric $d$ on the space of cdfs, we define the \emph{base sensitivity} of $d$ as the maximum distance between two adjacent ecdfs:
\[\Delta_d(n) = \sup_{\HH(x,x')\leq 1} d(\F_x,\F_{x'}).\]
 In what follows, we will consider several possible pseudo-metrics on cdfs, and show that  several pseudo-metrics of interest satisfy $\Delta_d(n)=1/n$. 


\subsection{Kolmogorov-Smirnov}
The Kolmogorov-Smirnov (KS) test is one of the most well known and commonly used distribution-free tests, and can be expressed in terms of the $L_\infty$-norm on cdfs:
\begin{equation}
    d_{KS}(F,G) = \sup_{t\in \mathbb{R}} |F(t)-G(t)|.
\end{equation}
Since $d_{KS}(F,G) = \lVert F-G\rVert_\infty$, it is a metric by the properties of norms. The base sensitivity of $d_{KS}$ is easily calculated, and first appeared in the proof of \citet[Lemma 5.1]{wasserman2010statistical}.

\begin{lemma}[\citealp{wasserman2010statistical}]
For any positive integer $n$, $\Delta_{d_{KS}}(n)=1/n$.
\end{lemma}

\subsection{Cram\'er-von Mises}
An alternative to Kolmogorov-Smirnov is Cram\'er-von Mises, which uses an $L_2$ norm, with respect to a base probability measure $H$:
\begin{equation}\label{eq:CvM}
d_{CvM}^H(F,G) = \left(\int_{-\infty}^\infty (F(t)-G(t))^2 \ dH(t)\right)^{1/2}.
\end{equation}
While in practice, $H$ is sometimes chosen based on the dataset, in this paper, we will assume that $H$ is a fixed probability measure that does not depend on the data. In the case of a goodness-of-fit test: $H_0:x_i\iid F$, it is natural to take $H=F$. 

If the distribution $H$ has support on the full real line (i.e., $H^{-1}(0,1)=\mathbb{R}$), then $d_{CvM}^H$ is a metric, since it is based on a norm. If this is not the case, then $d_{CvM}^H$ is a pseudo-metric, since if $F$ and $G$ only differ on a set of probability zero (with respect to $H$), then $d_{CvM}^H(F,G)=0$ even though $F\neq G$. 

In the following lemma, we show that the base sensitivity of $d_{CvM}^H$ is also $1/n$.
\begin{lemma}
For any positive integer $n$, and any probability measure $H$ on the space of cdfs,  $\Delta_{d_{CvM}^H}(n)=1/n$.
\end{lemma}

\begin{remark}
A variant of the Cram\'er-von Mises test is the Anderson-Darling test \citep{anderson1954test}. {\red While this test statistic can be expressed in terms of a pseudo-metric on cdfs, we show in Remark \ref{rem:anderson}, found in the appendix, that it does not have finite base sensitivity. Because of this, this test statistic cannot be directly used to achieve DP. This example illustrates that not all pseudo-metrics have finite base sensitivity, and we leave it to future researchers to explore alternative approaches to achieving DP in such settings.}
\end{remark}

\subsection{Weighted Wasserstein}
The Wasserstein metrics are related to optimal transport, representing how much mass must be moved to transform one distribution to another \citep{kantorovich1960mathematical,vaserstein1969markov}. For two cdfs $F$ and $G$, the Wasserstein metric of order $p$ is 
\[W_p(F,G) = \left( \int_{0}^1 |F^{-1}(u) - G^{-1}(u)|^p\ du\right)^{1/p},\]
and in the case that $p=1$, this simplifies to 
$W_1(F,G) = \int_{-\infty}^\infty |F(t)-G(t)| \ dt.$ 
For this paper, we modify the Wasserstein metric to include a base probability measure $H$:
\[d_W^H(F,G) = \int_{-\infty}^\infty |F(t) - G(t)| \ dH(t).\]
The purpose of the base measure $H$ is to ensure that the base sensitivity $d_W^H(\F_x,\F_{x'})$ is bounded for adjacent $x$ and $x'$. 

Just like the Cram\'er-von Mises metric, if $H$ has support on the full real line, then $d_W^H$ is a metric, whereas if this is not the case it is a pseudo-metric. 

\begin{lemma}
For any positive integer $n$, and any probability measure $H$ on the space of cdfs,  $\Delta_{d_{W}^H}(n) = 1/n$.
\end{lemma}

\subsection{Kuiper}
Another choice of metric is used in the Kuiper test:
\begin{equation}\label{eq:K}
    d_{K}(F,G) = \left(\sup_{t} F(t)-G(t)\right) + \left(\sup_{s} G(s)-F(s)\right),
\end{equation}
which is a variation on the KS distance. Note that rather than taking the largest absolute distance between $F(t)$ and $G(t)$, Kuiper's test considers both the supremum over $F(t)-G(t)$ as well as $G(t)-F(t)$. If either $F$ or $G$ stochastically dominates the other, then $d_K(F,G)=d_{KS}(F,G)$; but if this is not the case, such as with a difference in shape/scale then $d_K$ may be more sensitive than $d_{KS}$. 

Another aspect of $d_K$ is that it is also well defined for distributions on a circle, as it is invariant to cyclic transformations \citep{kuiper1960tests}. For example, this setting may occur if the random variables represent the time of year or time of day.

It is not immediately obvious that $d_K$ is a metric, so we establish this in the following lemma:
\begin{lemma}
The function $d_{K}(F,G) = \sup_{t} F(t)-G(t) + \sup_{s} G(s)-F(s)$ is a metric on the space of cdfs.
\end{lemma}

Finally, we derive the base sensitivity of $d_{K}$:
\begin{lemma}
For any positive integer $n$,  $\Delta_{d_{K}}(n)= 1/n$.
\end{lemma}

\section{Goodness-of-fit tests}\label{s:gof}
Suppose we want to test whether the observed data $x=(x_1,\ldots, x_n)$ were sampled i.i.d. from a distribution, $H_0: x_i \iid F$ or family of distributions, $H_0: x_i \iid F, F\in \mscr F$. For example, one may want to test if the data were drawn from $U(0,1)$ or from a parametric family such as $N(\mu,\sigma^2)$. Goodness-of-fit is an important class of hypothesis tests that is often used as a preliminary step in data analysis to verify whether model assumptions are reasonable. These tests can also be used to test the significance of an effect (e.g., $H_0: x_i \sim N(0,\sigma^2)$ for some $\sigma^2$).

\subsection{Goodness-of-fit methodology}

For a given pseudo-metric $d$ on cdfs, a natural test statistic for $H_0: x_i \iid F$ is $d(\F_x,F)$. This test statistic gives larger values the more that $\F_x$ differs from $F$. The most common test of this form is the Kolmogorov-Smirnov test,  however Cram\'er-von Mises, Wasserstein, and Kuiper's tests can also be applied. Under the null hypothesis $H_0:x_i\iid F$ for a continuous cdf $F$, it is easy to verify that the distribution $d(\F_x,F)$ does not depend on $F$ (although it does depend on $n$) for $d_{KS}$, $d_{K}$, $d_{CvM}^F$ and $d_{W}^F$. Note that for the Cram\'er-von Mises and Wasserstein metrics, we need to set $H=F$ in order for the sampling distribution to be independent of $F$. 

Sometimes instead of testing $H_0: x_i \iid F$ for a fixed and known $F$, we are interested in testing $H_0: x_i \iid F\in \mscr F$, where $\mscr F$ is a family of distributions (e.g., $N(\mu,\sigma^2)$). In this case, we propose using the test statistic $\inf_{G\in \mscr F} d(\F_x,G)$. Note that the calculation of this statistic requires finding the minimum distance estimates, which are generally different from maximum likelihood estimates (MLE). In the case of minimum KS estimates, there are efficient computational algorithms \citep{weber2006minimum}. In the case of Kolmogorov-Smirnov, tests of this form have been used to test normality in \citet{drezner2010modified}.

However, the sampling distributions of the quantities $\inf_{G\in \mscr F}d_{CvM}^H(\F_x,G)$ and $\inf_{G\in \mscr F}d_W^H(\F_x,G)$ will generally not be distribution-free. So, we limit our scope to $d_{KS}$ and $d_K$ when testing goodness-of-fit with unknown parameters.

Furthermore, even with $d_{KS}$ and $d_K$, it is not obvious that $\inf_{G\in \mscr F}d(\F_x, G)$ is distribution-free for arbitrary parametric families. We show in Proposition \ref{prop:distFree} that for location-scale families (e.g., $N(\mu, \sigma^2)$, $\mathrm{Laplace}(m,s)$, $\mathrm{Cauchy}(m,s)$, $\mathrm{Exp}(\lambda)$, etc.), both $\inf_{G\in \mscr F} d_{KS}(\F_x,G)$ and $\inf_{G\in \mscr F}d_K(\F_x,G)$ are distribution-free. Note that for these statistics the null distribution does depend on the family, but not on the unknown parameters. 

\begin{proposition}\label{prop:distFree}
Let $F_{0,1}$ be a continuous and invertible cdf and let 
\[\mscr F = \left\{F_{0,1}\left(\frac{\cdot-m}{s}\right)\middle| m\in \RR, s>0\right\},\]
be a location-scale family. Let  $x_i \iid F\in \mscr F$ for $i=1,\ldots,n$. Then both $\inf_{G\in \mscr F} d_{KS}(\F_x,G)$ and $\inf_{G\in \mscr F}d_K(\F_x,G)$ are  distribution-free for the null hypothesis $H_0: x_i \iid F\in \mscr F$. 
\end{proposition}

 To our knowledge, Proposition \ref{prop:distFree} is the first time it has been shown that the KS/Kuiper tests with unknown parameters are distribution-free, which is a useful result in its own right. As mentioned earlier, it is more common to estimate the unknown parameters using the MLE; however we are not aware of any results establishing that the test statistic is distribution-free when using MLEs.

In Theorem \ref{thm:good}, we derive the sensitivity for goodness-of-fit tests. Note that while the result is phrased in terms of a family of distributions, by taking the set $\mscr F$ to be a singleton, the known parameter result is a special case. 

\begin{theorem}\label{thm:good}
Let $(x_i)_{i=1}^n\in \RR^n$, and let $\mscr F$ be a family of cdfs. Let $d$ be a pseudo-metric on cdfs. Then, the statistic $T(x) = \inf_{G\in \mscr F} d(\F_x,G)$ has sensitivity $\Delta_d(n)$, with respect to adjacency metric $\HH$. Thus, $T(X)+\Delta_d(n) Z$ satisfies $\ep$-DP, where $Z$ is distributed as either $\mathrm{Tulap}(\exp(-\ep),0)$ or $\mathrm{Laplace}(1/\ep)$.
    \end{theorem}

\begin{remark}
For goodness-of-fit with known parameters, our sensitivity analysis of the KS test statistic is $1/n$, whereas \citet{wang2018statistical} derived a sensitivity of $2/n$. With our analysis, we are able to add half as much noise as \citet{wang2018statistical} while achieving the same privacy guarantee. 

\citet{wasserman2010statistical} derived the sensitivity of the KS statistic for use in the exponential mechanism in order to build a cdf estimator, and their sensitivity result agrees with Theorem \ref{thm:good} in the case of a simple null hypothesis; however, they did not consider privatizing the KS statistic for purposes of hypothesis testing, and did not consider other metrics.
\end{remark}

\subsection{Goodness-of-fit simulations} \label{s:goodSim}
In this section, we compare the performance of our proposed DP goodness-of-fit tests through simulations. All of the tests can be implemented in a straightforward manner, except for the Wasserstein and Cram\'er-von Mises tests. A computationally efficient implementation of the Cram\'er-von Mises test is included in Appendix \ref{app:good}. Through preliminary simulations, we found that the performance of the Wasserstein test was similar to Cram\'er-von Mises, yet was much more computationally expensive, so we omitted the Wasserstein test from the simulations presented here.

We first describe in detail the simulation setup; the simulations in the following sections are conducted in a similar manner.

A Monte Carlo simulation is used to experimentally perform power analysis. We measure the statistical power by repeatedly sampling the data from the true distribution and then calculating the privatized test statistics on the generated datasets. We set the significance level to $\alpha = 0.05$, and consider sample sizes $n\in \{50, 100, 200, 400, 800, 1600\}$. We consider $\epsilon=0.1$ and $\epsilon=1$; with larger $\epsilon$, the performance approaches that of the non-private tests. As all the hypothesis tests here are distribution-free, we approximate the null distribution of the privatized test statistics by simulating 1000 replicated samples under the normal distribution $N(0,1)$ and compute their test statistics for each value of $\ep$ and each sample size.

To implement the DP tests, we chose to use Tulap noise for the KS and Kuiper tests, and Laplace noise for the Cram\'er-von Mises test.

\begin{remark}
In this and the following simulation studies, we do not empirically measure the type I error of the candidate tests, since all candidates are proven to have exact type I error, up to Monto Carlo errors. Since the tests are calibrated using 1000 samples under the null distribution, all of the the type I errors are within standard error of $\pm 0.007$ of the nominal $.05$ level. For improved accuracy of type I errors, one can simply increase the number of Monte Carlo samples used to approximate the null distribution. {\red There are also techniques to derive conservative $p$-values from Monte Carlo simulations, such as in \citet{barber2022testing}.}
\end{remark}

\begin{figure}
    \centering
                    \vspace{-.5in}
    \includegraphics[width=.48\linewidth]{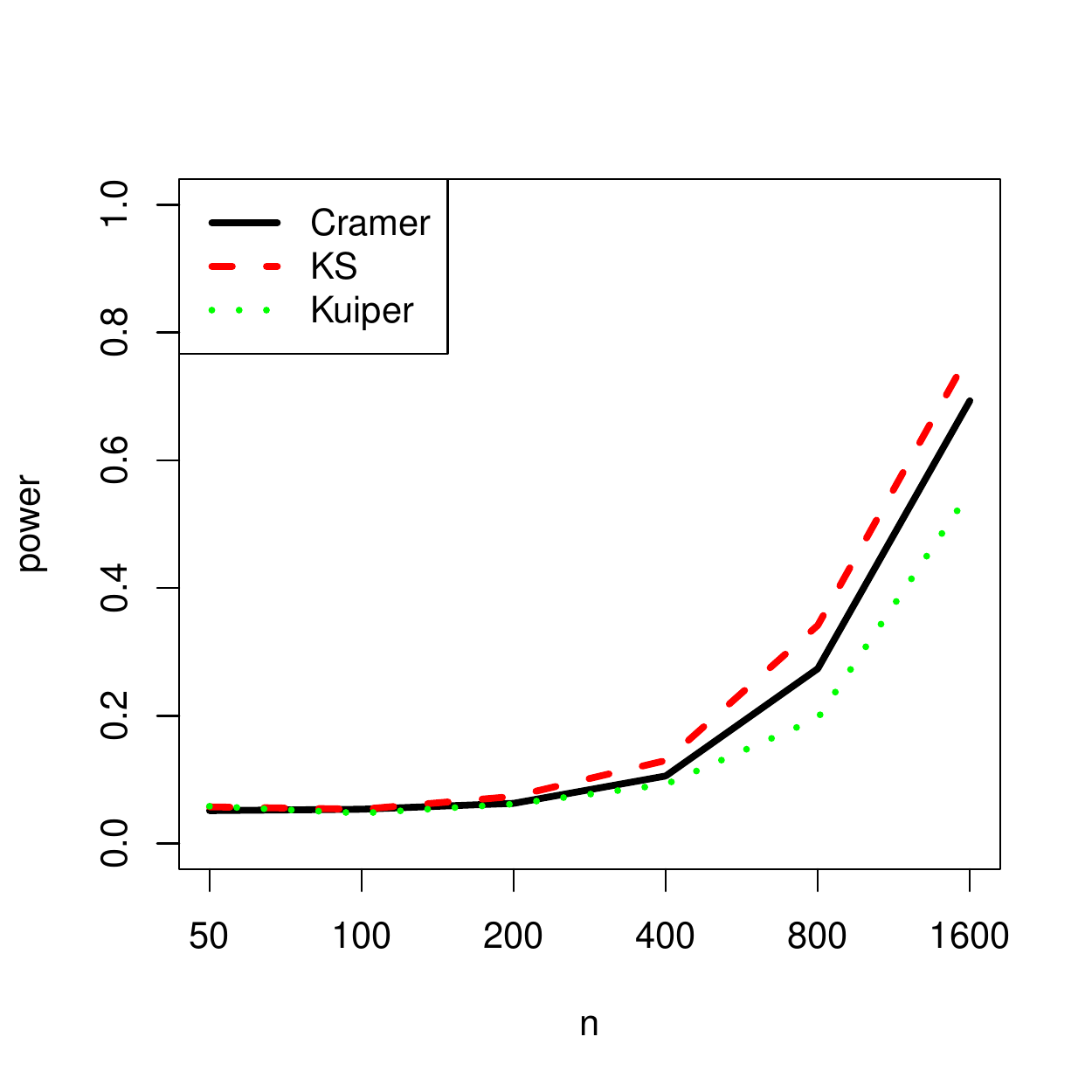}
        \includegraphics[width=.48\linewidth]{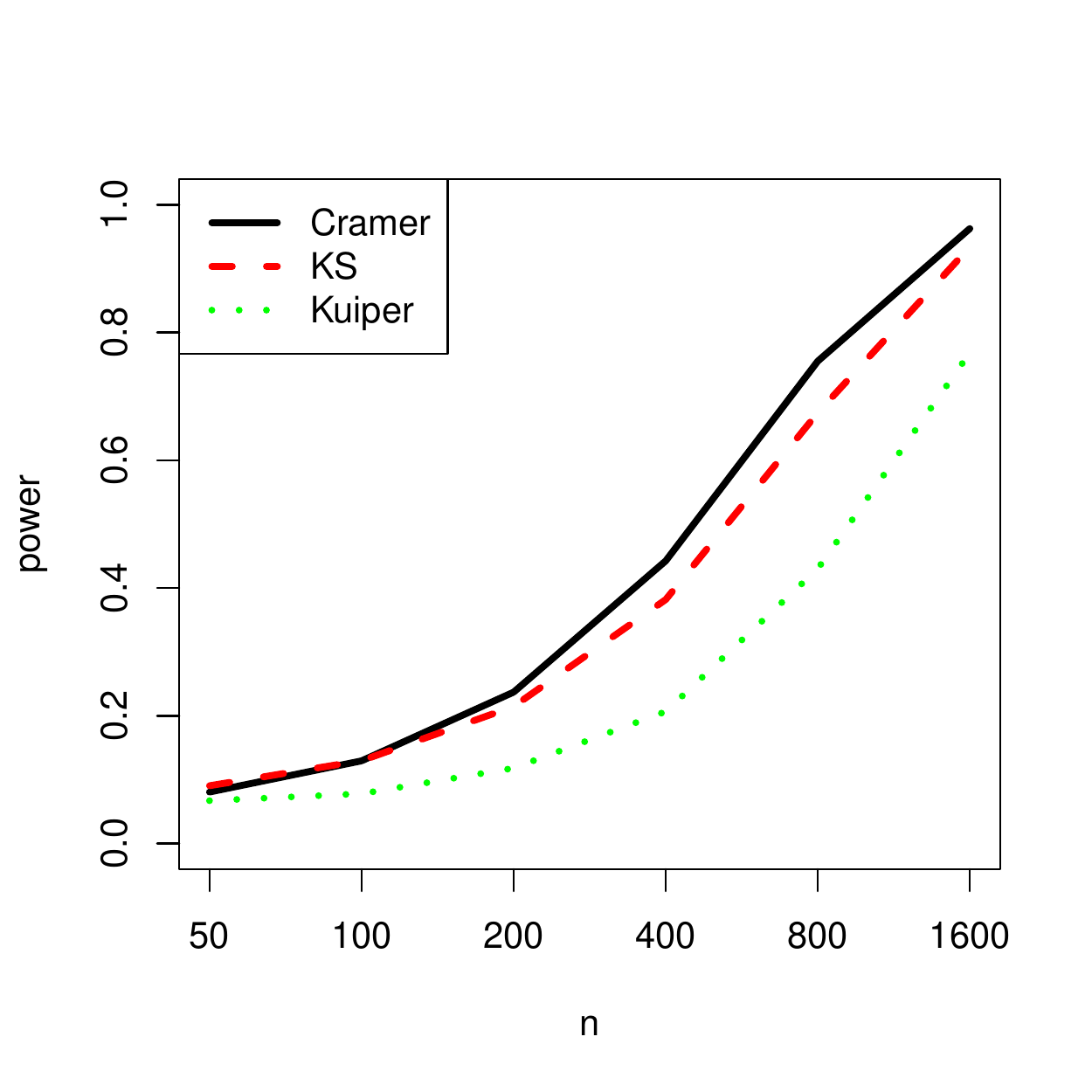}\\
                        \vspace{-.5in}
        \includegraphics[width=.48\linewidth]{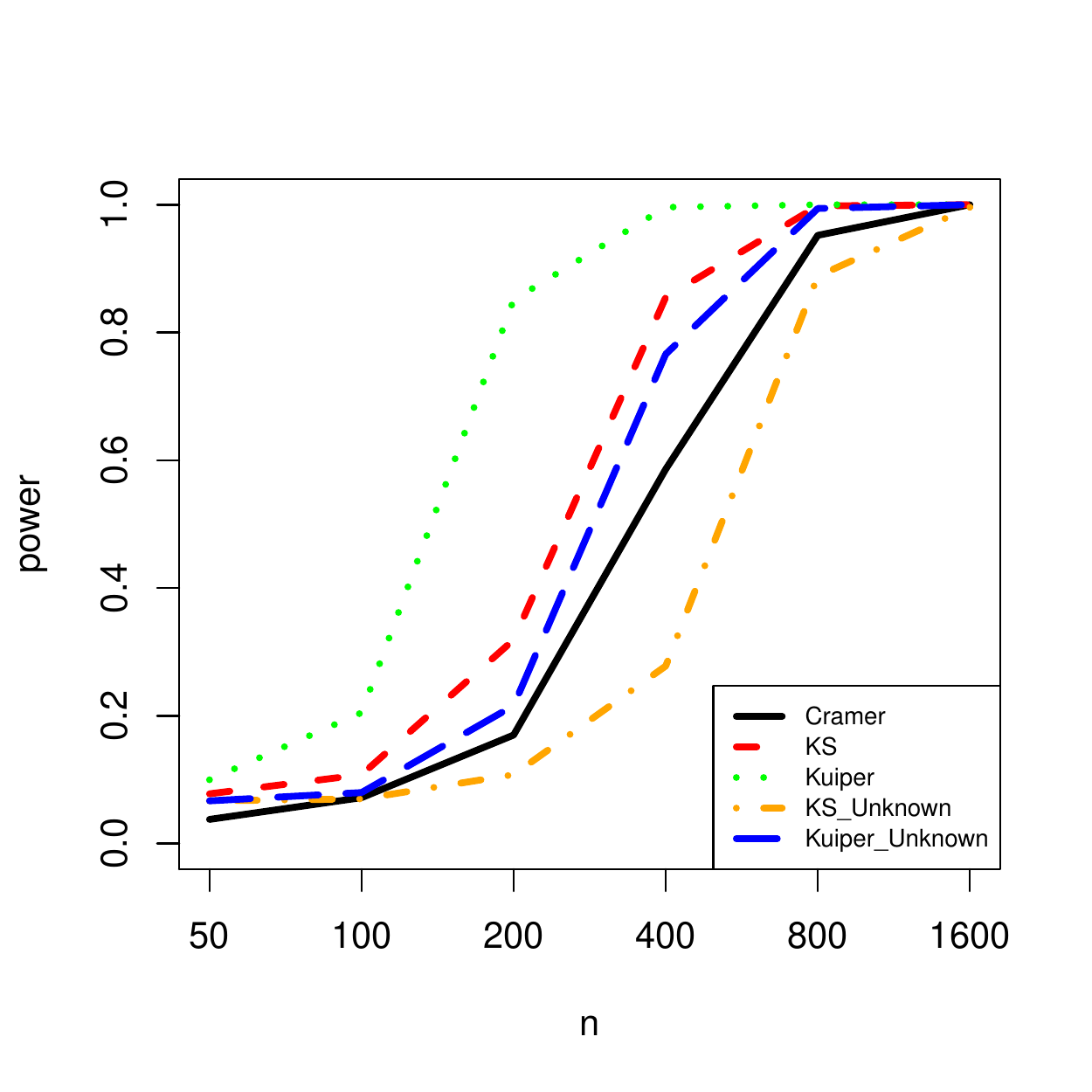}
        \includegraphics[width=.48\linewidth]{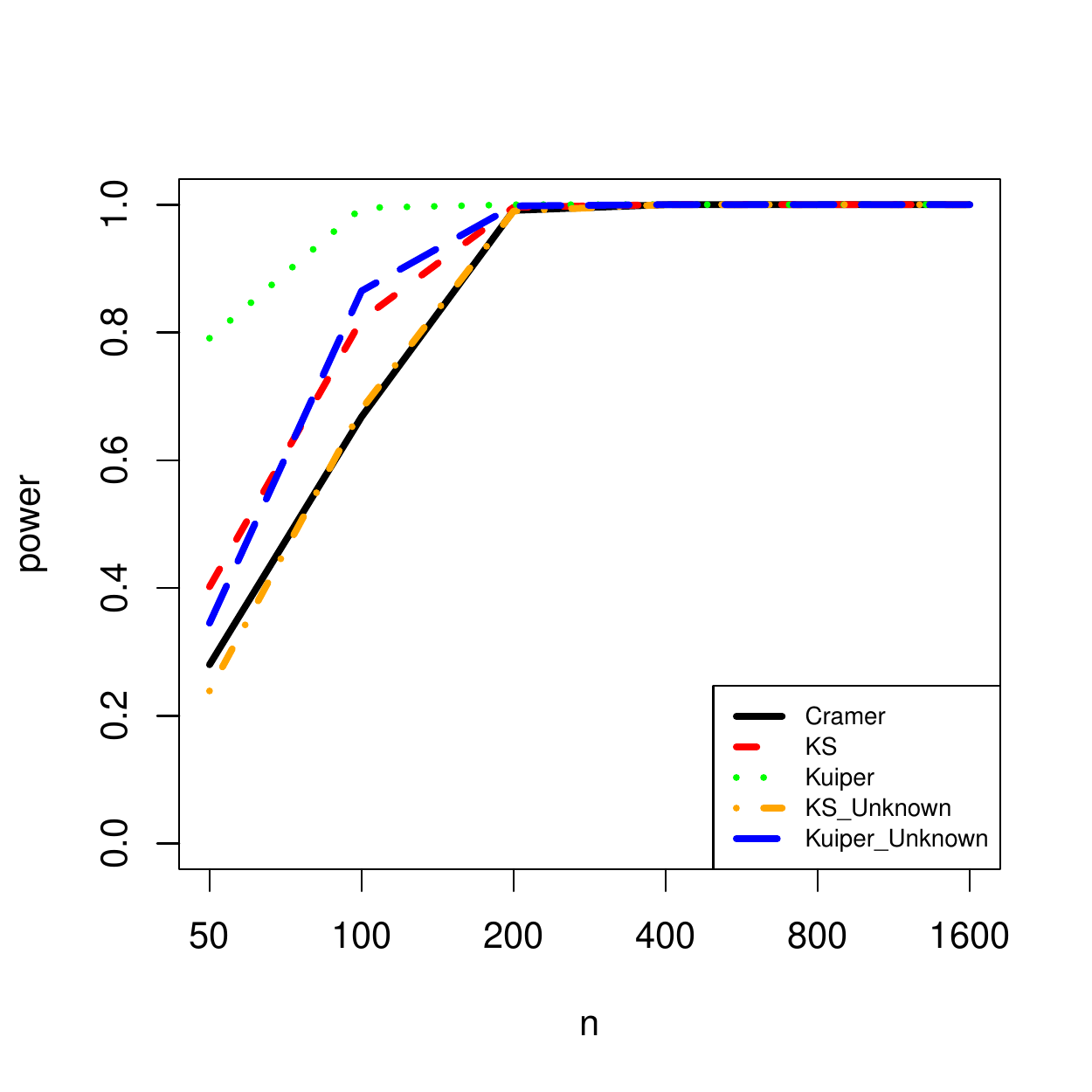}\\
                        \vspace{-.5in}
        \includegraphics[width=.48\linewidth]{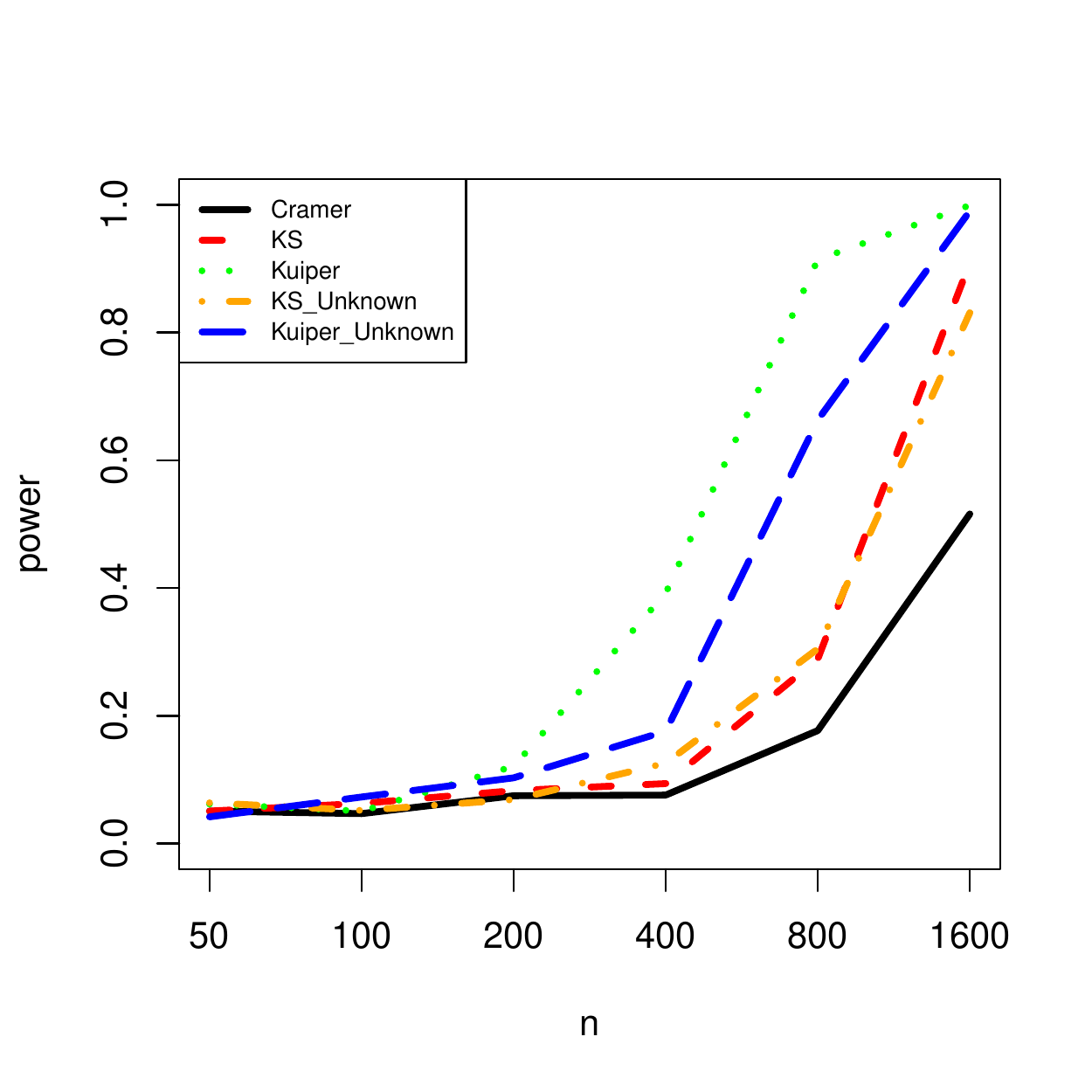}
        \includegraphics[width=.48\linewidth]{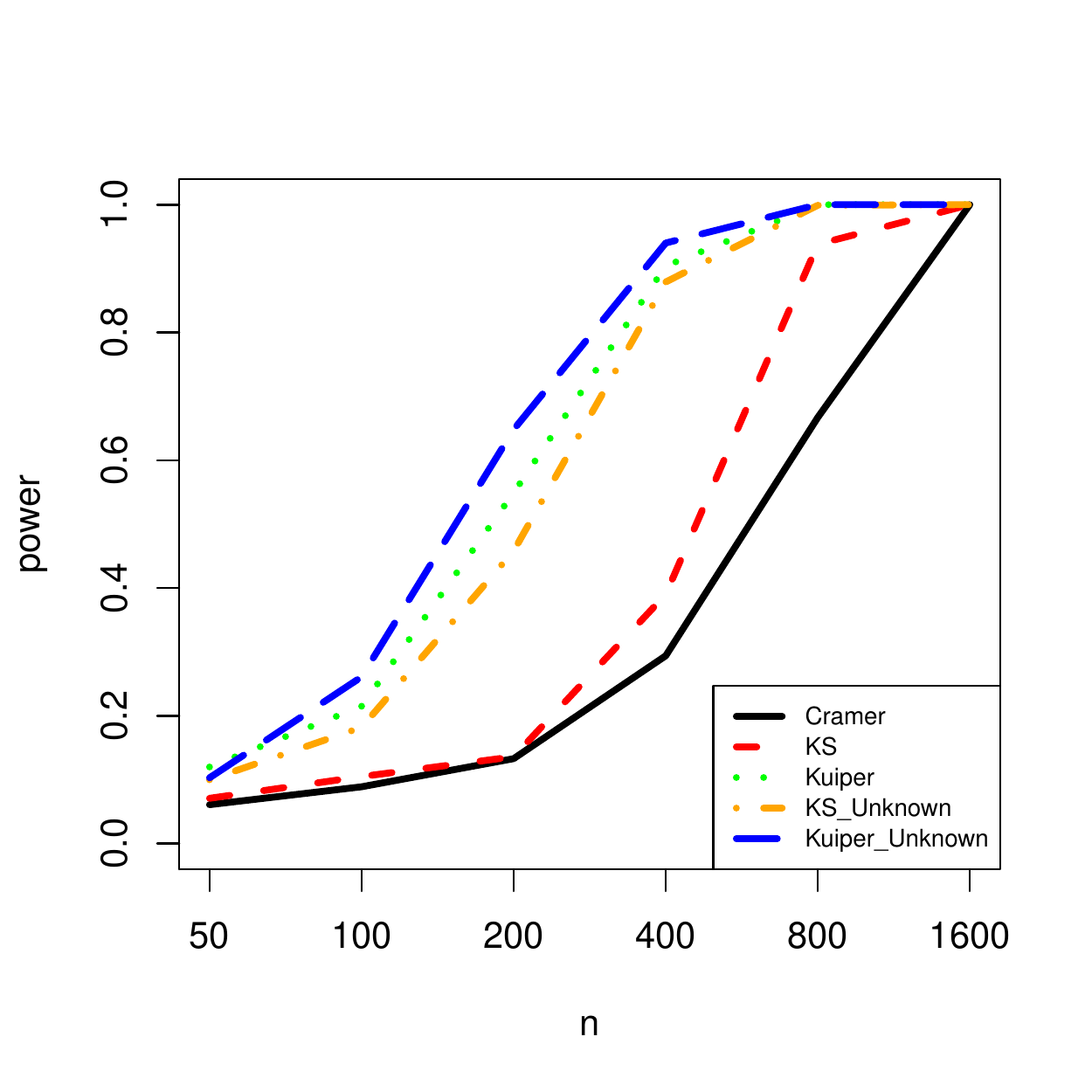}\\
        \vspace{-.25in}
    \caption{Goodness-of-fit. Top:  null distribution is $N(0,1)$, true distribution is $N(0.1,1)$. Middle: null distribution is $N(0,1)$, true distribution is $\mathrm{Cauchy}(0,1)$. Bottom: null distribution is $N(0,1)$, true distribution is $\mathrm{Laplace}(0,1)$. Left column: $\epsilon = 0.1$, right column: $\epsilon = 1$.}
    \label{fig:GOF}
\end{figure}


\noindent{\bf Known parameters: } Our null hypothesis is that the data are distributed as $N(0,1)$. We generate sample data from either $N(0.1,1)$, $\mathrm{Cauchy}(0,1)$ and $\mathrm{Laplace}(0,1)$ distributions. The results can be found in Figure \ref{fig:GOF}. Overall, we found that the KS and Cram\'er-von Mises tests were the best tests on the normally distributed data, with KS performing better when $\ep=0.1$ and Cram\'er-von Mises doing better with $\ep=1$. On the other hand, the Kuiper test performed better when the true data was more dispersed than the null hypothesis, significantly outperforming the KS and Cram\'er-von Mises tests. 

In general, we suspect that if the true distribution differs from the null distribution by a location-shift, that the KS test will be the better performer, whereas if there is also a difference in scale/shape the Kuiper test will be the stronger test. The Cram\'er-von Mises test seems to require lighter tailed distributions to perform well.

\noindent{\bf Unknown parameters: } We consider the KS and Kuiper tests when the parameters were unknown. To implement the tests, we estimate the mean and standard deviation of $N(\mu,\sigma^2)$ using the minimum KS and Kuiper distance. Then, we plug in the estimated parameters into null distribution and calculate test statistics between the generated sample dataset and null distribution. Finally, we add noise $N$ into the test statistics where $N\sim \mathrm{Tulap}(\exp(-\ep),0)$. 

We did not consider the KS and Kuiper tests with unknown parameters in the case of $N(0,1)$ versus $N(0.1,1)$, since the true distribution is part of the null hypothesis $H_0: N(\mu, \sigma^2)$. In the other simulation settings, we found that the Kuiper and KS tests with unknown parameters generally have inferior performance compared to the known parameter versions, which makes sense since they are operating with lesser information. Surprisingly, in the bottom right plot of Figure \ref{fig:GOF}, we found that the tests with unknown parameters actually outperformed the versions with known parameters. 
In all of the simulation settings considered, the Kuiper test with unknown parameters outperformed the KS test with unknown parameters. Our intuition is that with the null hypothesis of $N(\mu,\sigma^2)$, both the KS and Kuiper tests ignore differences in location and scale, leaving only differences in shape. As noted in the known parameter setting, the Kuiper test seems to be better than KS at detecting differences in shape.

\section{Two sample tests}\label{s:twosample}

In a two sample test, we have two independent samples of i.i.d. data, $x_1,\ldots, x_n$ and $y_1,\ldots,y_m$, and are interested in testing whether both samples came from the same distribution. The null hypothesis can be expressed as: $H_0: x_i\overset d=y_j$. A two sample test may be appropriate in randomized experiments, where the $x_i$ samples come from a control group and the $y_i$'s are from the treatment group. We may also be interested in doing a two sample test on observational data, such as men versus women, or smoking versus non-smoking. 

\subsection{Two sample methodology}
A natural test statistic for two sample tests is $d(\F_x,\F_y)$. Because there is no natural base measure in this setting, we will only consider the metrics $d_{KS}$ and $d_K$ for the test statistic. While there is a version of Cram\'er-von Mises which applies to two samples (and one could imagine a version of the Wasserstein metric as well), its base probability distribution is usually chosen to depend on the data, making our sensitivity analysis inapplicable. We leave it to future researchers to develop DP versions of these tests in the two sample case. It is easy to verify that the sampling distributions of $d_{KS}(\F_x,\F_y)$ and $d_K(\F_x,\F_y)$ are both distribution-free.

We assume that the values $n$ and $m$ are public knowledge, and do not require privacy protection. For $(x,y)$ and $(x',y')$ in $\RR^n\times \RR^m$, we will consider two versions of the Hamming metric: $\HH(x,x')+\HH( y,y')$ and $\max\{\HH(x,x'),\HH( y,y')\}$. In the first case, we consider $(x,y)$ and $(x',y')$ adjacent if either $\HH(x,x')= 1$ or $\HH( y,y')= 1$ (but not both), whereas in the second case  we consider $(x,y)$ and $(x',y')$ adjacent if both $\HH(x,x')\leq 1$  and $\HH( y,y')\leq 1$. The first case has been previously used in \citet{awan2018differentially}, \citet{awan2020one} and \citet{awan2021structure}, which allows one person to change their value, but not which group they belong to; the second case gives a stronger version of privacy, that would allow two individuals to switch groups as well as change their values. 


\begin{theorem}\label{thm:twosample}
Let $(x_i)_{i=1}^n$ and $(y_i)_{i=1}^m$ be two samples. Let $d$ be a pseudo-metric on cdfs. 
Then,
\begin{enumerate}
    \item  The statistic $T(x,y)=d(\F_x,\F_y)$ has sensitivity $\max\{\Delta_d(n),\Delta_d(m)\}$, with respect to the adjacency metric $m((x,y),(x',x'))=H(x,x')+H(y,y')$. Thus, $T(X)+\max\{\Delta_d(n),\Delta_d(m)\} Z$ satisfies $\ep$-DP, where $Z$ is distributed as either $\mathrm{Tulap}(\exp(-\ep),0)$ or $\mathrm{Laplace}(1/\ep)$.
    \item  The statistic $T(x,y)=d(\F_x,\F_y)$ has sensitivity $\Delta(n)+\Delta(m)$, with respect to the adjacency metric $m((x,y),(x',y'))= \max\{H(x,x'),H(y,y')\}$. Thus, $T(X)+[\Delta_d(n)+\Delta_d(m)] Z$ satisfies $\ep$-DP, where $Z$ is distributed as either $\mathrm{Tulap}(\exp(-\ep),0)$ or $\mathrm{Laplace}(1/\ep)$.
\end{enumerate}
\end{theorem}

Some alternative tests for $H_0: x_i \overset d= y_j$ are DP versions of the median test \citep{awan2018differentially}, absolute value Kruskal Wallis \citep{couch2019}, and Mann Whitney \citep{couch2019}. Note that the tests in \citet{couch2019} do not assume that the values $n$ and $m$ are publicly known, and because of this, the private Mann Whitney test from \citet{couch2019} satisfies $(\ep,\de)$-DP (a weaker notion of differential privacy), rather than $\ep$-DP. To make for a fairer comparison, we used the sensitivity calculation in \citet{couch2019} to design a modified private Mann-Whitney test, which uses the same assumptions as our proposed tests. The modified Mann-Whitney test is found in Appendix \ref{app:two}. As we will see in Section \ref{s:2simulations}, these prior tests are only able to capture particular differences in the distributions of $x$ and $y$ (such as a difference in median), whereas our DP KS and KP Kuiper tests can detect arbitrary differences in the distributions of $x$ and $y$. This may be a positive or negative, depending on whether one is interested in the broader null hypothesis $H_0: x_i \overset d=y_j$ or the narrower $H_0: \mathrm{median}(x)=\mathrm{median}(y)$. 

\subsection{Two sample simulations}\label{s:2simulations}
\begin{figure}
\centering
\vspace{-.5in}
  \includegraphics[width=0.48\linewidth]{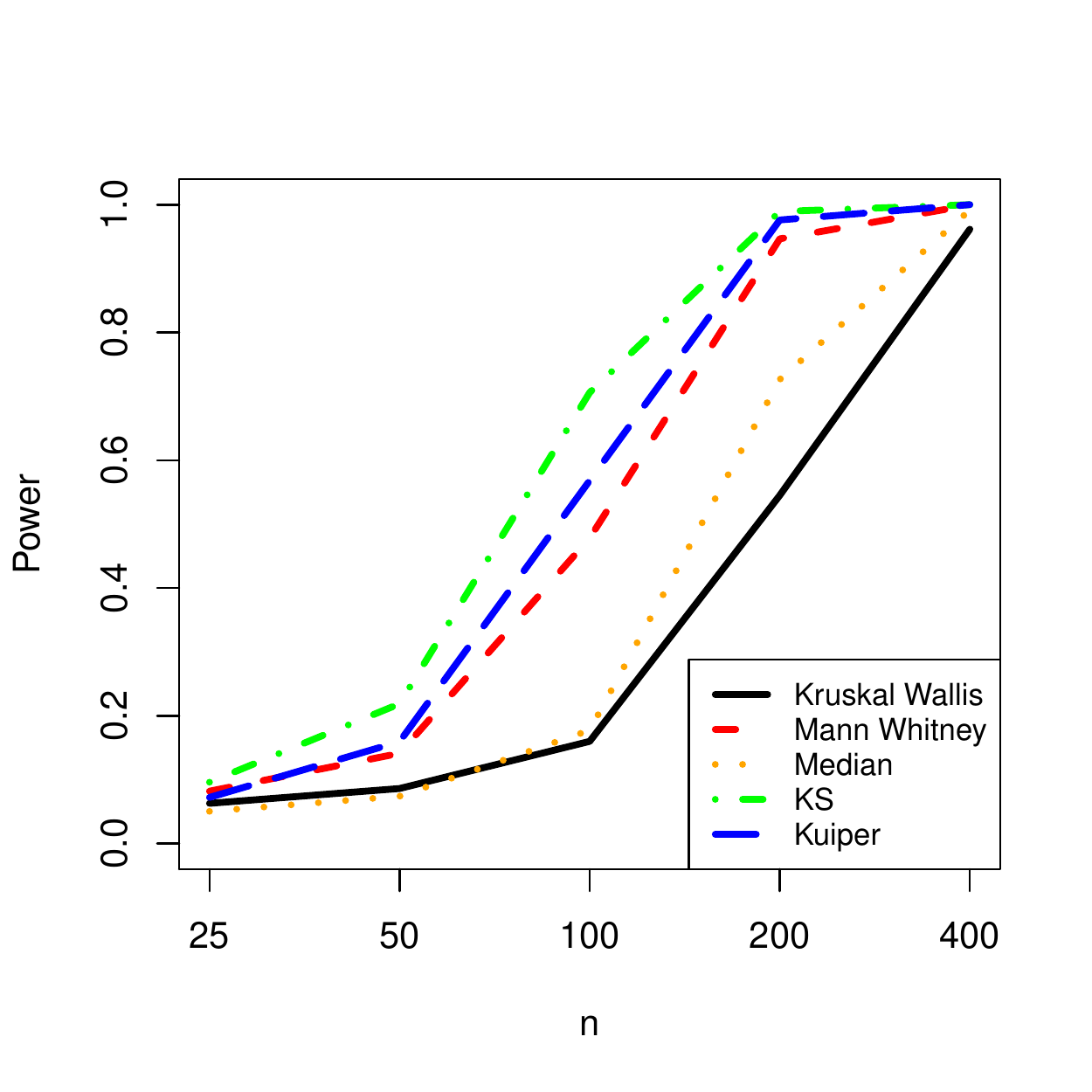}
  \includegraphics[width=0.48\linewidth]{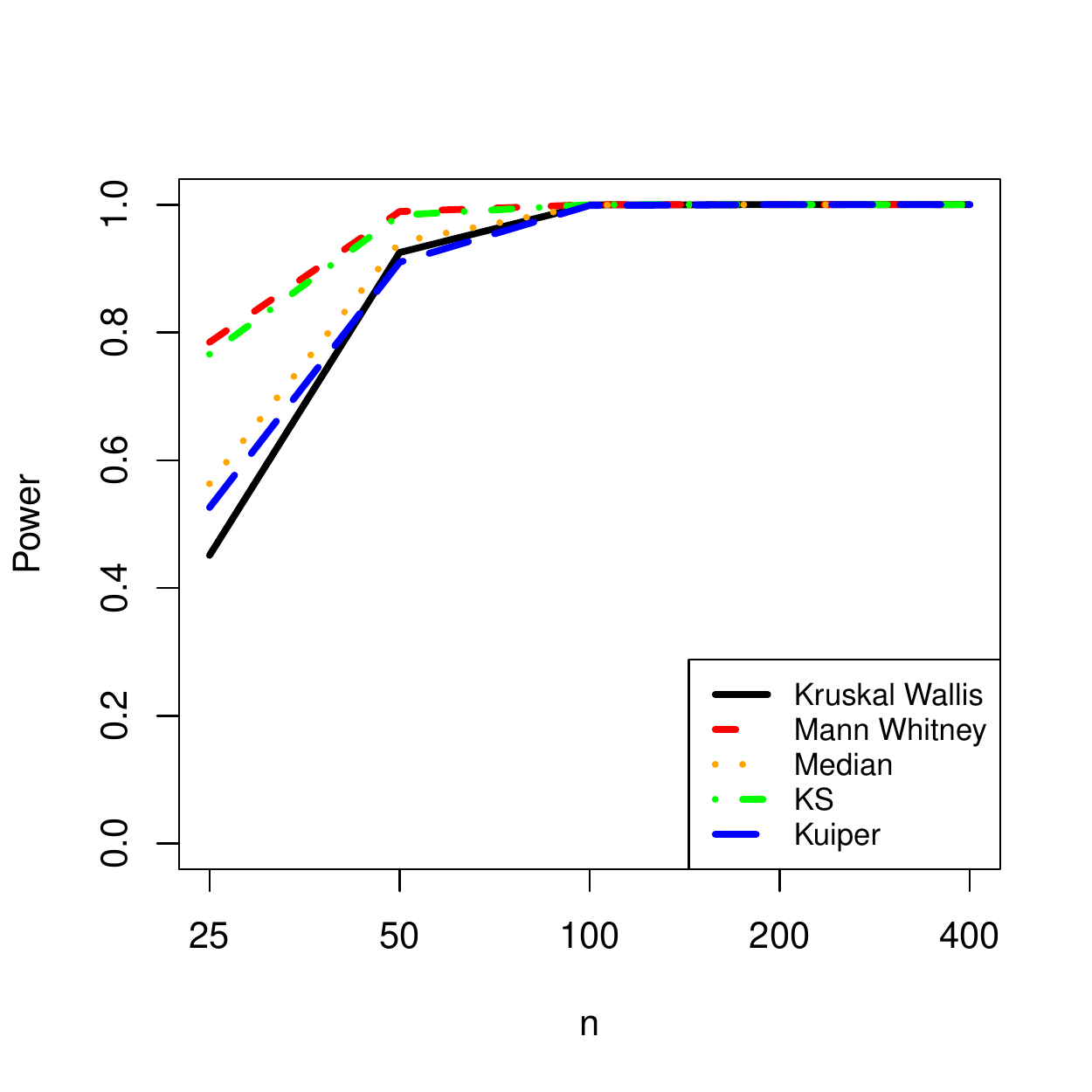}\\
    \vspace{-.5in}
    \includegraphics[width=0.48\linewidth]{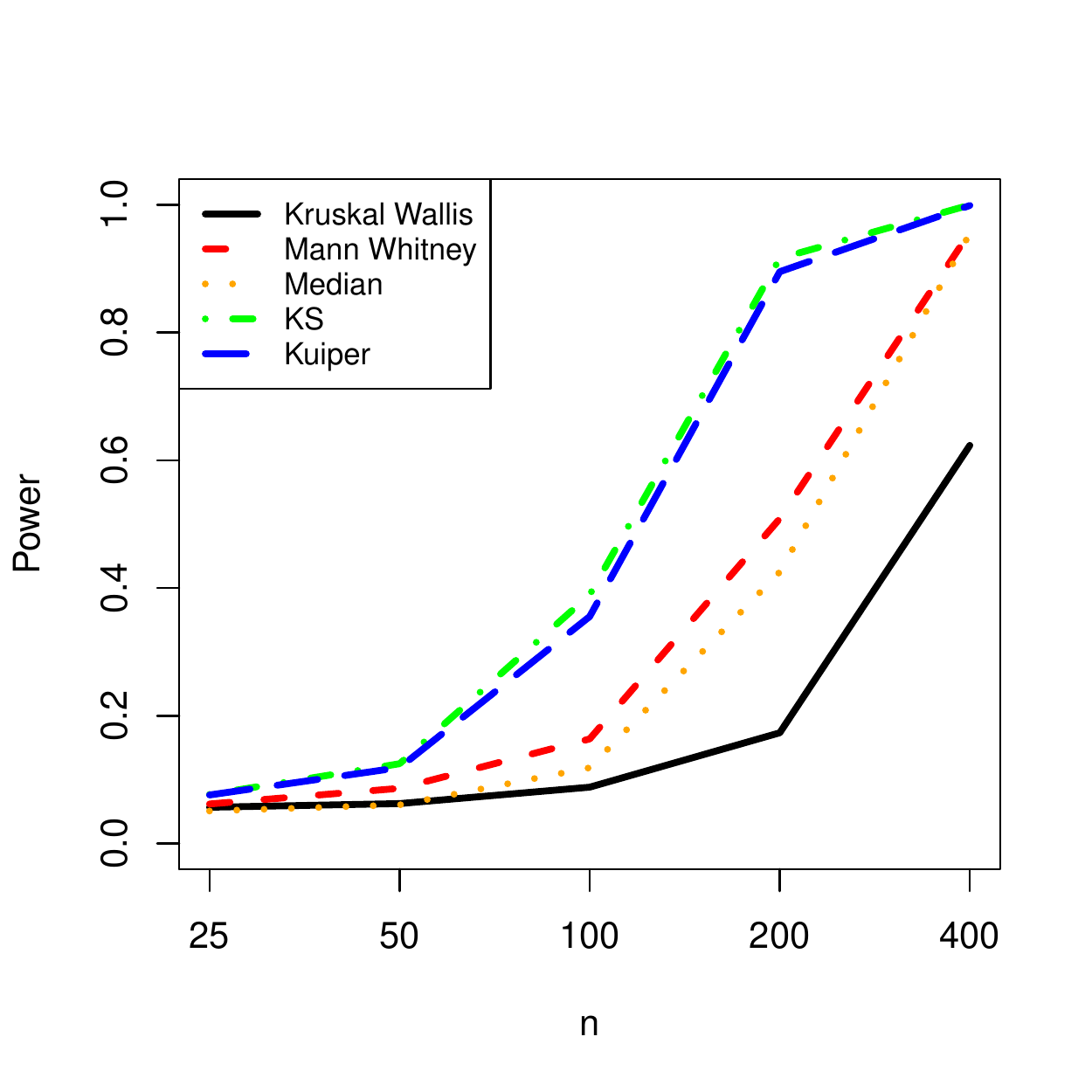}
  \includegraphics[width=0.48\linewidth]{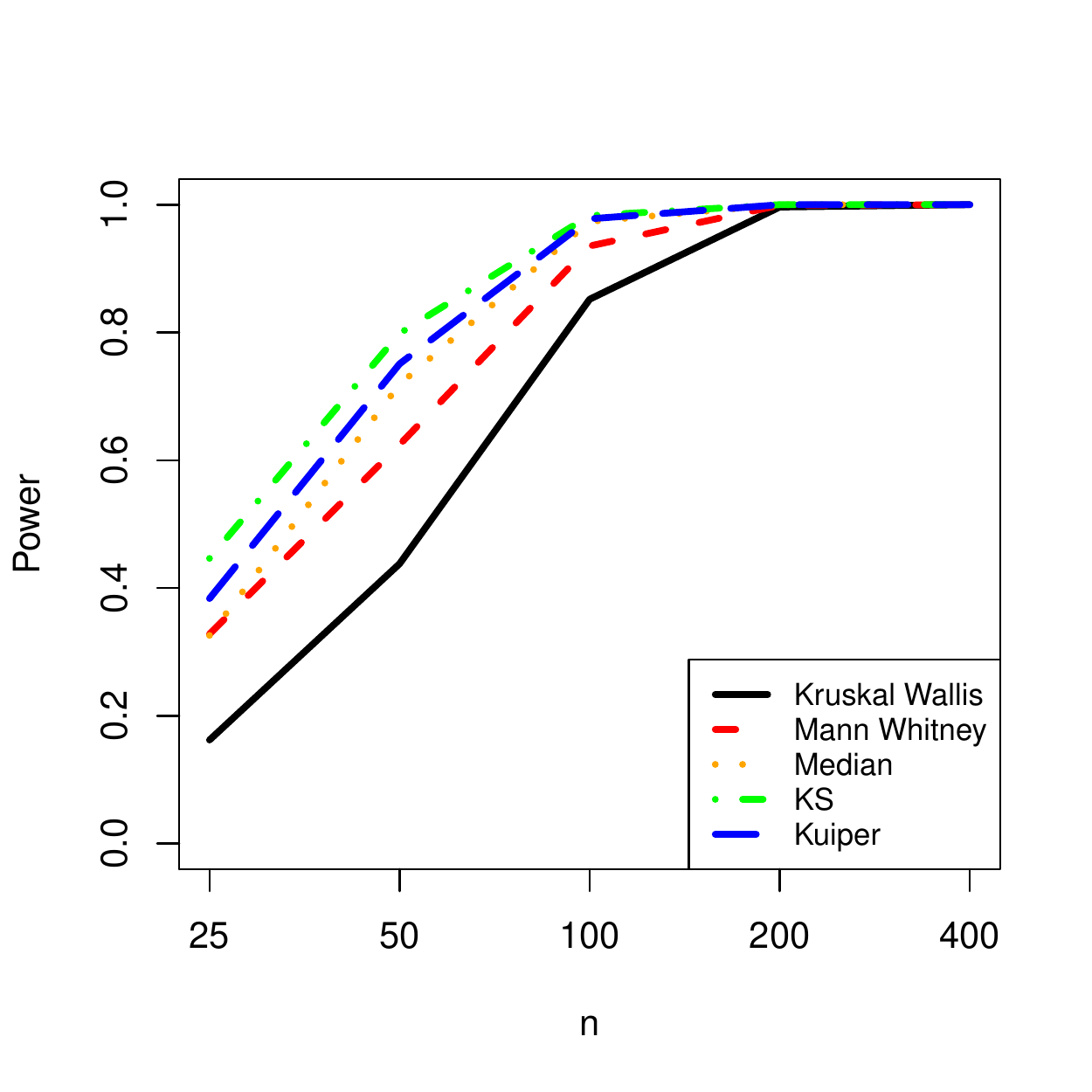}\\
  \vspace{-.5in}
  \includegraphics[width=0.48\linewidth]{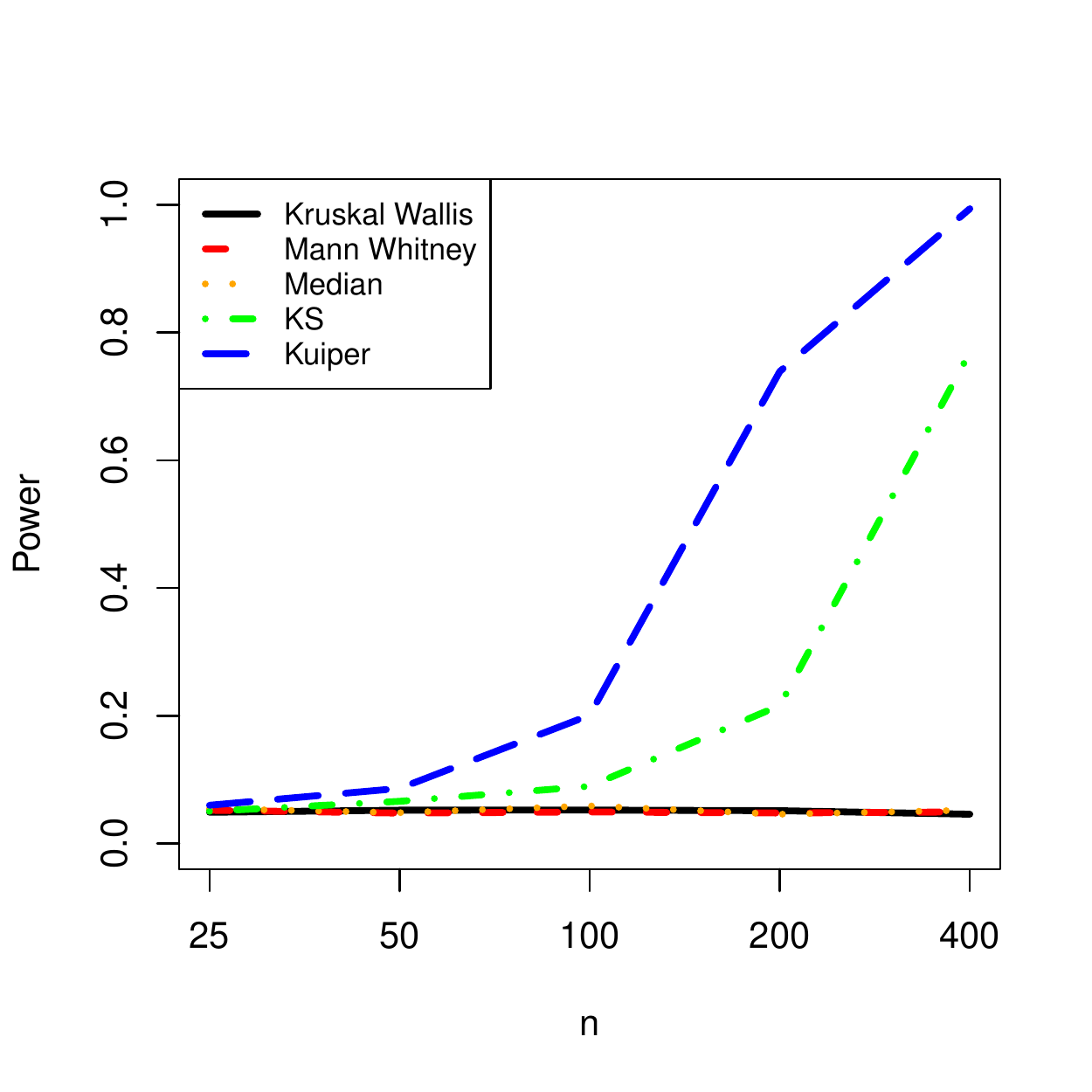}
  \includegraphics[width=0.48\linewidth]{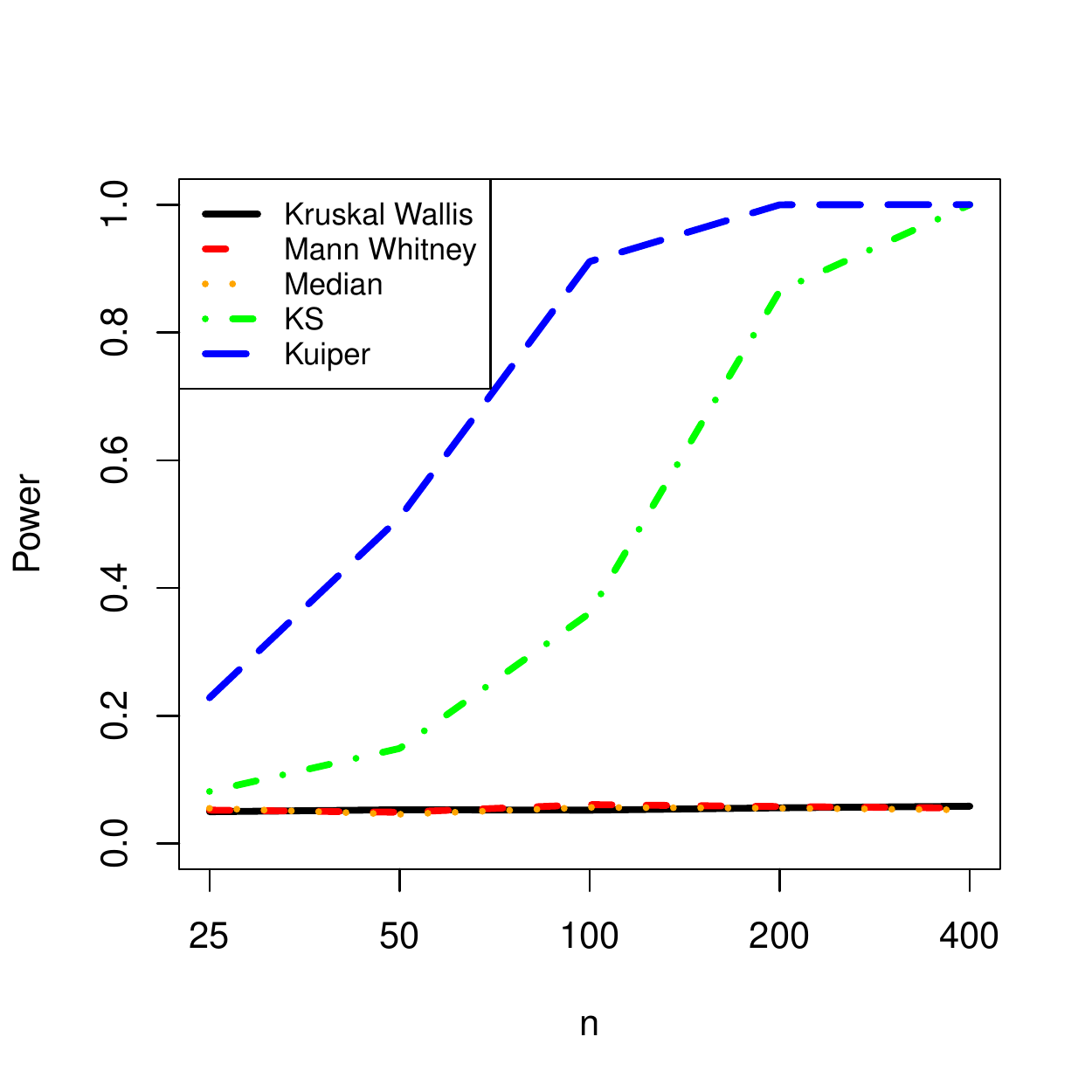}\\
          \vspace{-.25in}
\caption{Two sample. Top: $x\sim N(0,1)$ versus $y\sim N(1,1)$.  Middle: $x\sim \mathrm{Cauchy}(0,1)$ versus $y\sim \mathrm{Cauchy}(1,1)$. Bottom: $x\sim N(0,1)$ versus $y\sim \mathrm{Cauchy}(0,1)$. Left: $\epsilon = 0.1$, right: $\epsilon = 1$.}
\label{fig:two}
\end{figure}

Here we compare the statistical power of the DP two-sample tests. Algorithms for all of the competing tests are included in Appendix \ref{app:two}.  We use a similar simulation setup as Section \ref{s:goodSim} and set $m=n$. The simulation results are shown in Figure \ref{fig:two}. In the first and second lines of Figure \ref{fig:two}, we see that the KS test performs well when there is a location shift between $x$ and $y$. The Mann Whitney test is slightly better than the KS test in the setting of $N(0,1)$ versus $N(1,1)$ at $\ep=1$, but inferior in all other settings. The Kuiper test is comparable to the KS test in the case of $\mathrm{Cauchy}(0,1)$ versus $\mathrm{Cauchy}(1,1)$, but worse than KS when testing $N(0,1)$ versus $N(1,1)$. In the bottom row of Figure \ref{fig:two}, we are testing $N(0,1)$ versus $\mathrm{Cauchy}(0,1)$, and see that the Mann Whitney and Kruskal Wallis tests are not able to detect the difference in shape/scale in this case. In this setting, the Kuiper test is significantly more powerful than the KS test.

Similar to the goodness-of-fit simulation, we recommend the KS test for detecting changes in location, and the Kuiper test for detecting changes in shape/scale.

\section{Paired data tests}\label{s:paired}

Paired data may arise in a pre- versus post-treatment in an experiment. In this case, we observe $n$ i.i.d. pairs of values $(x_i,y_i)_{i=1}^n$, where $x_i$ and $y_i$ may have a non-negligible dependence. We want to test $H_0:x_i\overset d= y_i$. Often test statistics for this hypothesis operate  on the differences $z_i = y_i-x_i$, and either implicitly or explicitly have a broader null hypothesis. For example, the sign test has null hypothesis $H_0:\text{median}(x_i)=\text{median}(y_i)$. As such, all of the tests considered in this section (including prior tests) are not consistent against all possible alternatives. For example, if $x_i \sim N(0,1)$ and $y_i \sim N(0,2)$, all of the tests considered will have power equal to the type I error. 

\subsection{Paired data methodology}
In this section, we develop tests that operate under the null hypothesis $H_0:[(y_i-x_i)\text{ is symmetric about zero}]$. Using the notation $z_i = y_i-x_i$, our test statistic is $d(\F_z,\F_{-z})$. Similar to the two-sample case, there is no natural base measure in this setting, so, we focus on $d_{KS}$ and $d_K$.

\begin{corollary}\label{cor:symmetry}
Let $(z_i)_{i=1}^n\in \RR^n$. Let $d$ be a pseudo-metric on cdfs. 
Then, the statistic $T(z) = d(\F_z,\F_{-z})$ has sensitivity $2\Delta(n)$, with respect to the adjacency metric $\HH$, where $\F_{-z}$ is the ecdf of $(-z_i)_{i=1}^n$.  Thus, $T(X)+2\Delta_{d}(n) Z$ satisfies $\ep$-DP, where $Z$ is distributed as either $\mathrm{Tulap}(\exp(-\ep),0)$ or $\mathrm{Laplace}(1/\ep)$.
\end{corollary}

Alternative DP tests for $H_0:x_i\overset d= y_i$ include the sign test \citep{awan2018differentially}, and the Wilcoxon signed rank test \citep{couch2019}. 

\begin{remark}
The DP Kolmogorov-Smirnov test of \citet{yu2018differentially} also operates on paired-data, but is specifically designed for the setting where one datapoint is the true observed value, and the other is a predicted value, from some model. The sensitivity calculation of \citet{yu2018differentially} for their DP KS test also gave a value of $2/n$, which is a special case of part 2 of Theorem \ref{thm:twosample}.
\end{remark}

\begin{remark}
In the non-private case, the KS statistic has higher Bahadur efficiency than the sign test \citep{chatterjee1973kolmogorov}, and in turn the Kuiper test has higher Bahadur efficiency than the KS test \citep{littell1974relative}. Nevertheless, these are asymptotic results, and the finite sample performance may vary. In Section \ref{s:pairedSimulation}, we find that the performance of the privatized tests depends on the true distribution as well as the privacy budget $\ep$. 
\end{remark}

\begin{remark}
   {\red While the tests developed in this section are designed for paired data hypothesis tests, they can also be applied to test whether the distribution of univariate data $z_i$ is symmetric about zero.}
\end{remark}

\subsection{Paired data simulations}\label{s:pairedSimulation}
 \begin{figure}
    \centering
    \vspace{-.5in}
    \includegraphics[width=.48\linewidth]{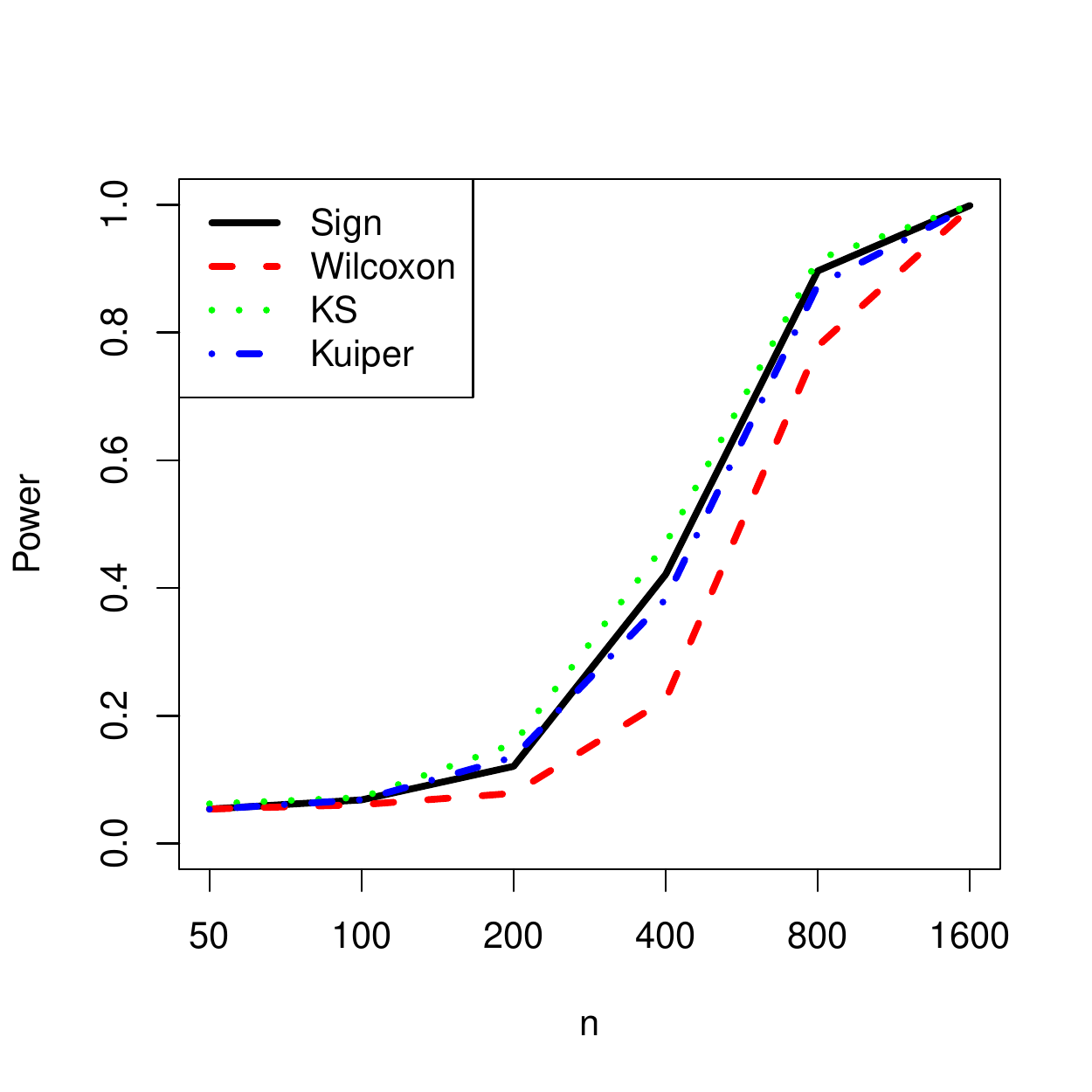}
        \includegraphics[width=.48\linewidth]{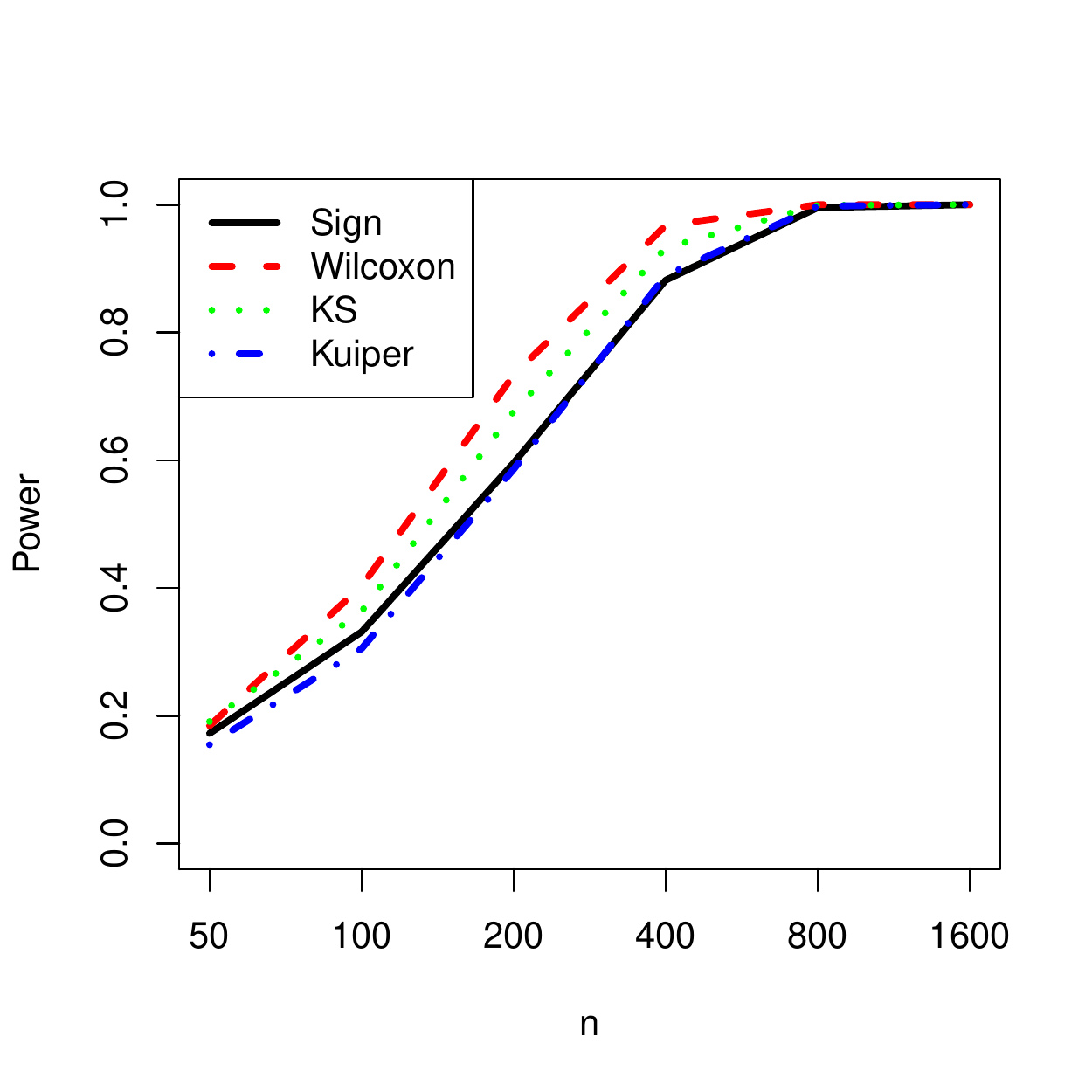}\\
        \vspace{-.5in}
 \includegraphics[width=.48\linewidth]{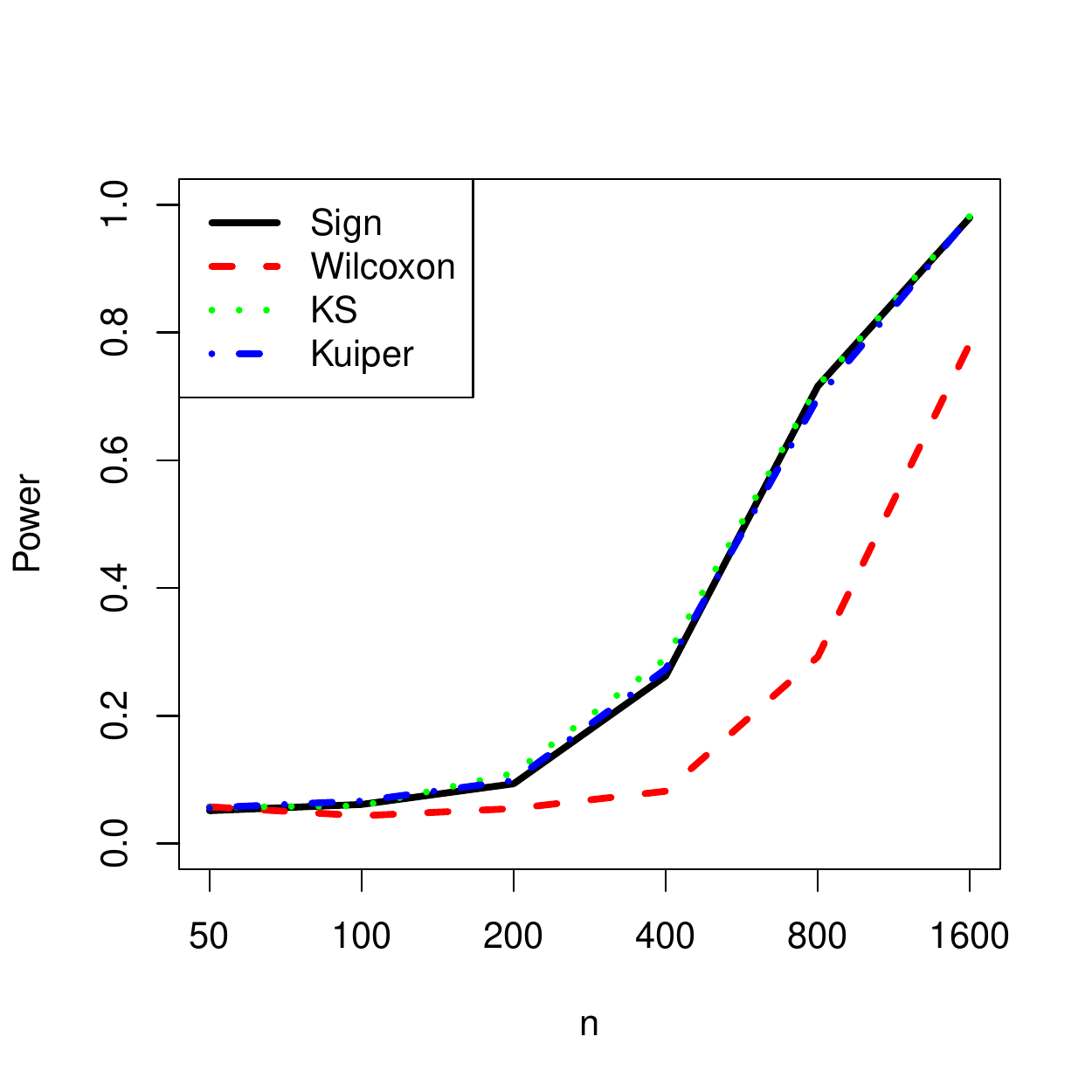}
        \includegraphics[width=.48\linewidth]{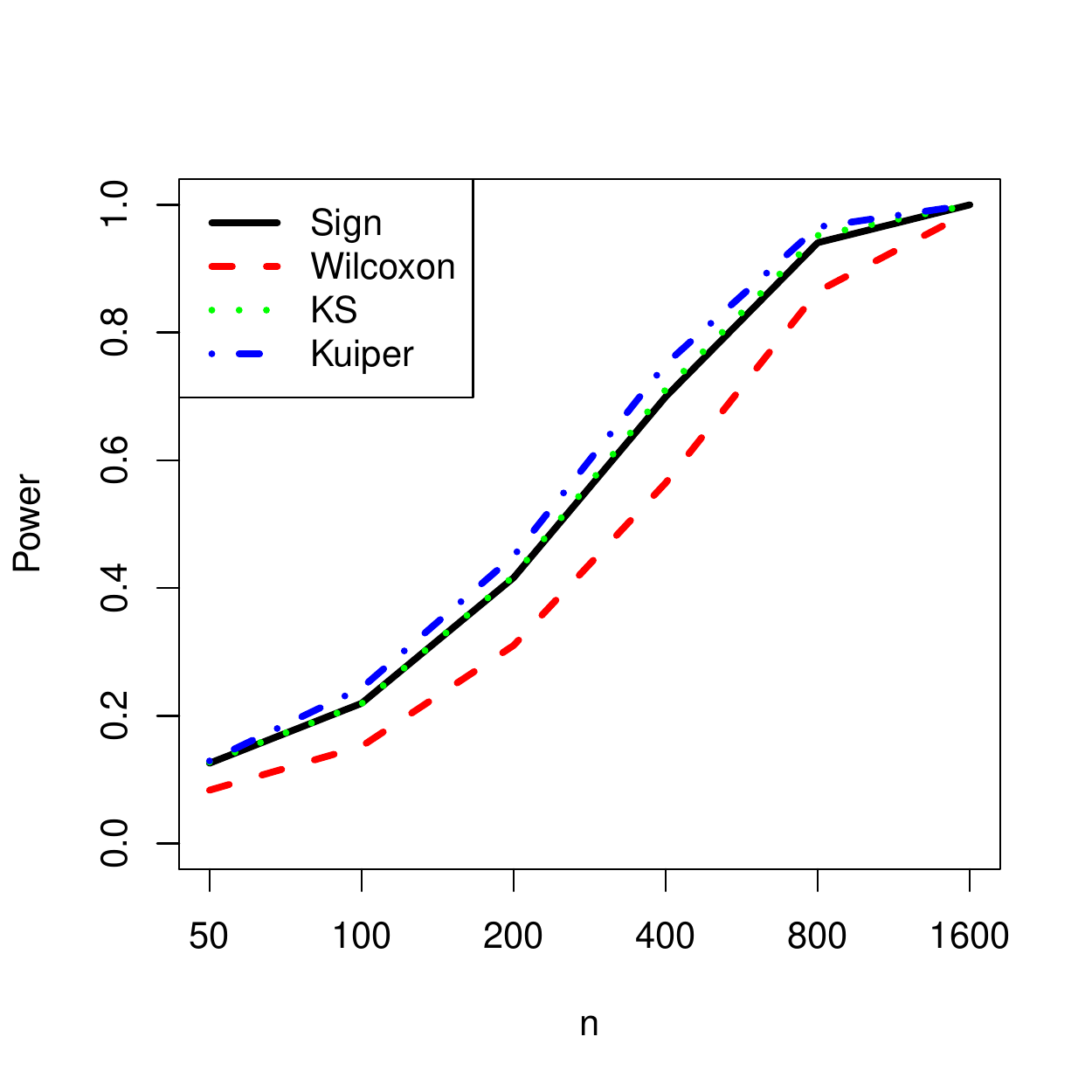}\\
    \vspace{-.5in}
 \includegraphics[width=.48\linewidth]{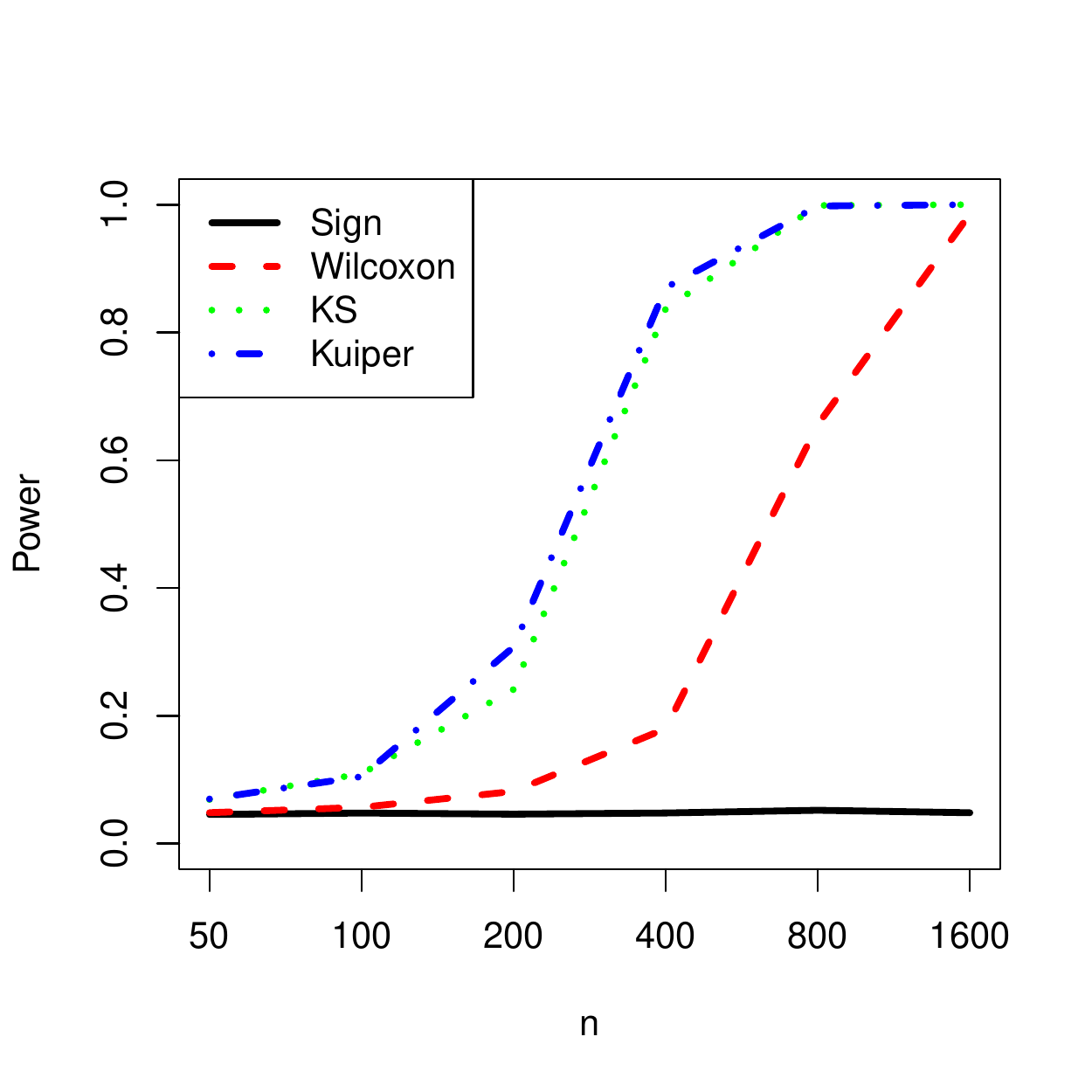}
        \includegraphics[width=.48\linewidth]{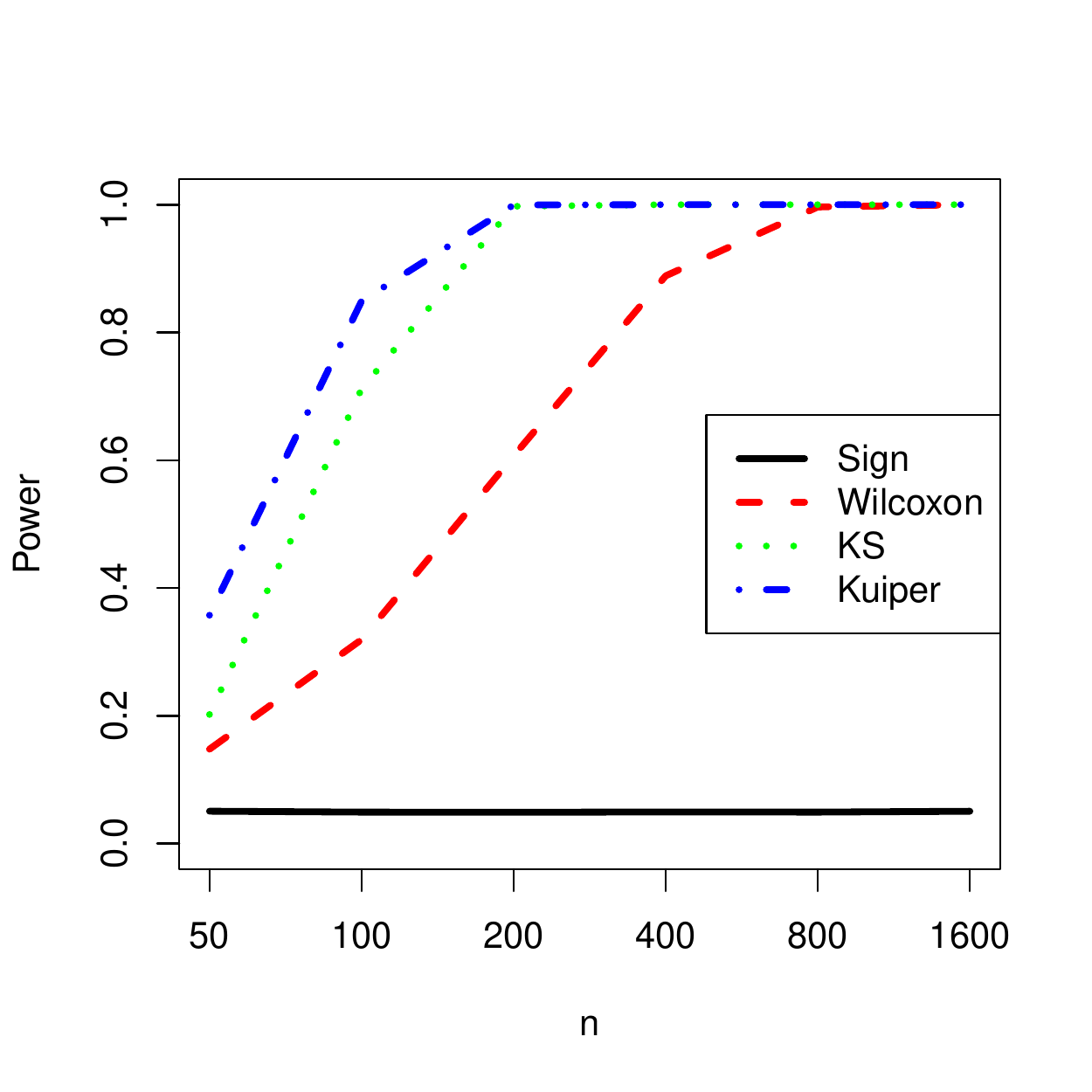}\\
                \vspace{-.25in}
    \caption{Paired-data. Top: $y - x \sim N(0.2,1)$. Middle: $y-x\sim \mathrm{Cauchy}(0.2,1)$. Bottom: $y-x+\log(2)\sim \mathrm{Exp}(1)$. Left: $\epsilon = 0.1$, Right: $\epsilon = 1$.}
    \label{fig:paired}
\end{figure}

In this section, we compare the private KS and Kuiper tests against the private sign test and Wilcoxon signed rank test for paired data, through simulations. Algorithms for the private sign test and Wilcoxon signed rank tests can be found in Appendix \ref{app:paired}. The simulation setup is similar to Section \ref{s:goodSim}. The simulation results are shown in Figure \ref{fig:paired}.

We found that the Wilcoxon signed rank test was the most powerful test when $y-x$ is normally distributed and $\ep=1$, however, in all other settings, both the KS and Kuiper tests outperformed  Wilcoxon. The sign test performed surprisingly well in both the normal and Cauchy settings, but had no power when the median of the data was 0 (see bottom row of Figure \ref{fig:paired}), {\red which is to be expected, since it operates on the broader null hypothesis $H_0: \mathrm{median}(y_i-x_i)=0$}. We found that the KS test outperformed the Kuiper and sign tests for normally distributed data, whereas the Kuiper test did best with the Cauchy and exponential data. 

Overall, the KS and Kuiper tests perform well in all settings considered, with KS being stronger with normally distributed data, and Kuiper being better in the other settings. If one is interested in the null hypothesis $H_0: \mathrm{median}(y_i-x_i)=0$ rather than $H_0: [(y_i-x_i) \text{ is symmetric about zero}]$,  then the sign test is preferred, as it avoids false positives due to a change in shape/scale. Wilcoxon is only recommended with normally distributed data and moderate to large $\ep$.

\section{Discussion}\label{s:discussion}

 
 In this paper, we showed that several test statistics based on ecdfs have low sensitivity enabling private hypothesis tests with minimal noise. In particular, we showed that the pseudo-metric structure along with the notion of base sensitivity made sensitivity calculations of the test statistics very straightforward. We developed several new DP hypothesis tests, 
 {\red achieve state-of-the-art performance in settings with either small $\ep$ values or heavy-tailed data, and are still competitive with normally distributed data. }

While the DP tests in this paper are designed to satisfy $\ep$-DP, they are easily modified to satisfy a variety of DP guarantees. For example, to satisfy $(\ep,\de)$-DP, one can replace $\mathrm{Tulap}(\exp(-\ep),0)$ with $\mathrm{Tulap}(\exp(-\ep),\frac{2\delta b}{1-b+2\delta b})$ \citep{awan2018differentially}. More generally, to satisfy $f$-DP \citep{dong2022gaussian} (a generalization of DP, phrased in terms of hypothesis testing), we can replace $\mathrm{Tulap}(\exp(-\ep),0)$ with a canonical noise distribution \citep{awan2023canonical}. There are also additive noise mechanisms for divergence-base definitions of DP \citep{bun2016concentrated,bun2018composable}.

As all of the DP hypothesis tests developed in this paper add independent noise to a test statistic, the asymptotic framework of \citet{wang2018statistical} can be used to develop approximate sampling distributions for our tests. The result of the framework is a convolution of the asymptotic distribution of the non-private test statistic (such as the Kolmogorov distribution in the KS test) and the distribution of the privacy noise (e.g., Laplace or Tulap). We chose not to explore this direction in this paper, since Monte Carlo approximations are easily implemented, and incorporate the sample size into the sampling distribution approximation. However, for larger sample sizes, the asymptotic approximations of \citet{wang2018statistical} could be both highly accurate and computationally efficient. 

{\red The notion of base sensitivity, proposed in this paper,  can also be used to facilitate the sensitivity calculation for the exponential mechanism \citep{mcsherry2007mechanism}, when the loss function is of the form $d(\hat \theta(x),\theta)$ for a pseudo-metric $d$ on a space $\Theta$. Indeed the sensitivity of this loss function is $\sup_{\theta} \sup_{H(x,x')\leq 1}|d(\hat\theta(x),\theta)-d(\hat\theta(x'),\theta)|\leq \sup_{H(x,x')} d(\hat\theta(x),\hat\theta(x'))$, where the right hand side is the base sensitivity of $d$ (similar to Theorem \ref{thm:good}). For example, base sensitivity was used in this way to produce differentially private persistence diagrams, a topological data analysis method, via the exponential mechanism \citep{kang2023differentially}.}


\appendix

\section{Proofs}\label{app:proofs}

\begin{proof}[Proof of Lemma 1]
Let $x,x'\in \RR^n$ such that $H(x,x')\leq 1$. Without loss of generality, suppose that $x$ and $x'$ differ in the first entry. Then,
  \begin{align*}
        d_{KS}(\F_x,\F_{x'})&= \sup_t \left|\F_x(t)-\F_{x'}(t)\right|\\
        &=\sup_t \left| \frac 1n \sum_{i=1}^n I(x_i\leq t)-\frac 1n \sum_{i=1}^n I(x_i'\leq t)\right|\\
        &=\sup_t \left|\frac 1n I(x_1\leq t)-\frac 1n I(x_1'\leq t)\right|\\
        &\leq \frac 1n,
    \end{align*}
    since $0\leq |I(x_1\leq t)-I(x_1'\leq t)|\leq 1$ for all $t$. This shows that $\Delta_{d_{KS}}(n)\leq 1/n$. To see that the upper bound it tight, note that if $x$ and $x'$ are not identical, then $d_{KS}(\F_x,\F_{x'})\geq 1/n$.
\end{proof}

\begin{proof}[Proof of Lemma 2]
Let $x,x'\in \RR^n$ such that $H(x,x')\leq 1$. Without loss of generality, suppose that $x$ and $x'$ differ in the first entry. Then,
 \begin{align*}
        d_{CvM}^H(\F_x,\F_{x'})&=\left( \int_{-\infty}^\infty (\F_x(t)-\F_{x'}(t))^2 \ dH(t) \right)^{1/2}\\
        &=\left( \int_{-\infty}^\infty \left[\frac 1n I(x_1\leq t)-\frac 1n I(x_1'\leq t)\right]^2 \ dH(t) \right)^{1/2}\\
        &\leq \left( \int_{-\infty}^\infty \frac {1}{n^2} \ dH(t) \right)^{1/2}\\
        &=\frac 1n,
    \end{align*}
    where we use the fact that $\int_{-\infty}^\infty \ dH(t)=1$ and that $[I(x_1\leq t)-I(x_1'\leq t)]\leq 1$. This shows that $\Delta_{d_{CvM}^H}(n)\leq 1/n$. To see that the bound is tight, suppose that $x_1$ is arbitrarily small, and $x_1'$ is arbitrarily large. In the limit, we have $d_{CvM}^H(\F_x,\F_{x'})\rightarrow1/n$, giving a matching lower bound on $\Delta_{d_{CvM}^H}(n)$.
\end{proof}


\begin{remark}\label{rem:anderson}
A variant of the Cram\'er-von Mises test is the Anderson-Darling test \citep{anderson1954test}, which is based on the following pseudo-metric:

\[d_{And}^H(F,G) = \left(\int_{-\infty}^\infty \frac{(F(t)-G(t))^2}{H(t)(1-H(t))} \ dH(t)\right)^{1/2}.\]

When used for testing goodness-of-fit, $H$ is usually set to $F$ and the test statistic is of the form $d_{And}^F(\F_x,F)$. While $d_{And}^F$ is a pseudo-metric, it does not have finite base sensitivity: 
\begin{align*}
    d_{And}^F(\F_x,\F_{x'})&\leq \frac 1n \left(\int_{-\infty}^\infty \frac{1}{F(t)(1-F(t))} \ dF(t) \right)^{1/2} \\
    &= \frac 1n \left(\int_{0}^1 \frac{1}{x(1-x)} \ dx\right)^{-1/2}\\
    &=\infty.
\end{align*} 
Because of this, this test statistic cannot be directly used to achieve DP. 
\end{remark}

\begin{proof}[Proof of Lemma 3]
Let $x,x'\in \RR^n$ such that $H(x,x')\leq 1$. Without loss of generality, suppose that $x$ and $x'$ differ in the first entry. Then,
\begin{align*}
    d_{W}^{H}(\F_x,\F_{x'})&= \int_{-\infty}^\infty |\F_x(t)-\F_{x'}(t)| \ dH(t)\\
    &= \int_{-\infty}^\infty \left|\frac 1n I(x_1\leq t)-\frac 1n I(x_1'\leq t)\right| \ dH(t)\\
    &\leq \int_{-\infty}^\infty \frac 1n \ dH(t)\\
    &=\frac 1n,
\end{align*}
 where we use the fact that $\int_{-\infty}^\infty \ dH(t)=1$. To see that the bound is tight, suppose that $x_1$ is arbitrarily small, and $x_1'$ is arbitrarily large. In the limit, we have $d_{W}^H(\F_x,\F_{x'})\rightarrow1/n$, giving a matching lower bound on $\Delta_{d_{W}^H}(n)$.
\end{proof}

\begin{proof}[Proof of Lemma 4]
First note that $d_K(F,G)=0$ if and only if $F=G$, and that $d$ is symmetric in the roles of $F$ and $G$. It remains to show that the triangle inequality holds. Let $F,G$, and $H$ be three cdfs. Then,

\begin{align*}
    d_K(F,G)&= \sup_{t} F(t)-G(t) + \sup_{s} G(s)-F(s)\\
    &=\sup_{t} F(t)-H(t)+H(t)-G(t) \\
    &\phantom{=}+ \sup_{s} G(s)-H(s)+H(s)-F(s)\\
    &\leq \sup_{t_1} F(t_1)-H(t_1) + \sup_{t_2} H(t_2)-G(t_2)\\
    &\phantom{=}+\sup_{s_1} G(s_1)-H(s_1) + \sup_{s_2} H(s_2)-F(s_2)\\
    &= \sup_{t_1} F(t_1)-H(t_1)  + \sup_{s_2} H(s_2)-F(s_2)\\
    &\phantom{=}+ \sup_{t_2} H(t_2)-G(t_2) +\sup_{s_1} G(s_1)-H(s_1)\\
    &=d_K(F,H)+d_K(H,G).
\end{align*}
\end{proof}

\begin{proof}[Proof of Lemma 5]
Let $x,x'\in \RR^n$ such that $H(x,x')\leq 1$. Without loss of generality, suppose that $x$ and $x'$ differ in the first entry. Then,
 \begin{align*}
       & d_{K}(\F_x,\F_{x'})= \sup_{t} \F_x(t)-\F_{x'}(t) + \sup_{s} \F_{x'}(s)-\F_x(s)\\
        &=\sup_t \left[\frac 1n I(x_1\leq t)-\frac 1n I(x_1'\leq t)\right] + \sup_s \left[\frac 1n I(x_1'\leq s)-\frac 1n I(x_1\leq s)\right].
    \end{align*}
    At this point either $x_1 < x_1'$, in which case the first supremum achieves the value of $1/n$ and the second supremum achieves the value of $0$; or $x_1'> x_1$, in which case the first supremum achieves the value of $0$ and the second seupremum achieves the value of $1/n$. In either case, we have that $d_{K}(\F_x,\F_{x'})= 1/n$, establishing that $\Delta_{d_{K}}(n)= 1/n$. 
\end{proof}

\begin{proof}[Proof of Proposition 3]
We present the proof for $d_{KS}$; the proof for $d_K$ is similar. We will denote by $F_{m,s}$ the cdf in $\mscr F$ with location $m$ and scale $s$. Then $F_{m,s}(t) = F_{0,1}\left(\frac{t-m}{s}\right)$ and $F_{m,s}^{-1}(u) = sF^{-1}_{0,1}(u)+m$. 

Suppose that  $x_i\iid F_{m,s}$. Define $u_i =F_{m,s}(x_i)$, which are distributed $u_i\iid U(0,1)$, and satisfy $x_i=F^{-1}_{m,s}(u_i)$. Then,
\begin{align}
    \inf_{m^*,s^*} \sup_{t\in \RR}&\left|\frac 1n \sum_{i=1}^n I(x_i\leq t)-F_{m^*,s^*}(t)\right|\\
    &=\inf_{m^*,s^*} \sup_{t\in \RR}\left| \frac 1n \sum_{i=1}^n I(F_{m,s}^{-1}(u_i)\leq t)-F_{m^*,s^*}(t)\right|\\
    &=\inf _{m^*,s^*} \sup_{t\in \RR}\left| \frac 1n \sum_{i=1}^n I(u_i\leq F_{m,s}(t))-F_{m^*,s^*}(t)\right|\label{eq:free1}\\
     &=\inf _{m^*,s^*} \sup_{t\in \RR}\left| \frac 1n \sum_{i=1}^n I(u_i\leq F_{m,s}(t))-F_{m^*,s^*}\left(F_{m,s}^{-1}\left(F_{m,s}(t)\right)\right)\right|\label{eq:free2}\\
    &=\inf _{m^*,s^*} \sup_{v\in [0,1]}\left| \frac 1n \sum_{i=1}^n I(u_i\leq v) - F_{m^*,s^*}\left(F^{-1}_{m,s}(v)\right)\right|\label{eq:free3}\\
    &=\inf_{\tilde m,\tilde s} \sup_{v\in [0,1]} \left|\frac 1n \sum_{i=1}^n I(u_i\leq v)-F_{0,1}\left( \frac{F_{0,1}^{-1}(v)-\tilde m}{\tilde s}\right)\right|\label{eq:free4},
\end{align}
where \eqref{eq:free1} and \eqref{eq:free2} use the invertibility of $F_{m,s}$ and
\eqref{eq:free3} uses the fact that $F_{m,s}$ is a continuous cdf and so has $(0,1)\subset F^{-1}_{m,s}(\RR)\subset[0,1]$; \eqref{eq:free4} used the following change of variables: $\tilde m=\frac{m^*s^*}{s}-\frac ms$, and $\tilde s = \frac {s^*}{s}$, where $(m^*,s^*)\mapsto(\tilde m,\tilde s)$ is a bijection. We see that the last formulation does not depend on the original $m$ and $s$. 
\end{proof}

   \begin{proof}[Proof of Theorem 1]
    Let $x,x'\in \RR^n$ such that $\HH(x,x')\leq 1$. Then 
    \begin{align*}
        \inf_{G\in \mscr F} d(\F_x,G)-\inf_{H\in \mscr F} d(\F_{x'},H)
        &\leq \inf_{G\in \mscr F} d(\F_x,\F_{x'})+d(\F_{x'},G)-\inf_{H\in \mscr F}d(\F_{x'},H)\\
        &=d(\F_x,\F_{x'})\\
        &\leq \Delta_d(n).
    \end{align*}
    Swapping the roles of $x$ and $x'$ we have that 
    \[\left|\inf_{G\in \mscr F} d(\F_x,G)-\inf_{H\in \mscr F} d(\F_{x'},H)\right|\leq \Delta_d(n).\]
    \end{proof}

\begin{proof}[Proof of Theorem 2]
\begin{enumerate}
    \item Suppose that one entry of $y$ changes. Then the adjacent databases are $(x,y)$ and $(x,y')$, where $x\in \RR^n$ and $y,y'\in \RR^m$, where $\HH(y,y')\leq 1$. Then, 
    \[|T(x,y)-T(x,y')|=|d(\F_{x},\F_{y})-d(\F_x,\F_{y'})|\leq d(\F_y,\F_{y'})\leq \Delta_d(m).\]
    If instead we compare the samples $(x,y)$ and $(x',y)$, similar calculations give $\Delta_d(n)$.
    \item Let $(x,y),(x',y')\in \RR^n\times \RR^m$ such that $\HH(x,x')\leq 1$ and $\HH(y,y')\leq 1$. Then, 
    \begin{align*}
        |T(x,y)-T(x',y')|&=|d(\F_{x},\F_{y})-d(\F_{x'},\F_{y'})|\\
        &=|d(\F_{x},\F_{y})- d(\F_{x},\F_{y'}) + d(\F_{x},\F_{y'})-   d(\F_{x'},\F_{y'})|\\
        &\leq |d(\F_{x},\F_{y})- d(\F_{x},\F_{y'})| + |d(\F_{x},\F_{y'})-   d(\F_{x'},\F_{y'})|\\
        &\leq d(\F_y,\F_{y'})+d(\F_x,\F_{x'})\\
        &\leq \Delta_d(m)+\Delta_d(n).
    \end{align*}
\end{enumerate}
\end{proof}

\begin{proof}[Proof of Corollary 3]
 Let $\HH(z,z')\leq 1$. Notice that $z$ and $z'$ differ in at most one entry, and $-z$ and $-z'$ also differ in at most one entry. We see that Corollary 3 is a special case of part 2 of Theorem 2, where $m=n$. 
\end{proof}

\section{Test algorithms}

In this appendix, we give pseudo-code for tests included in the simulations. The simulation code is available at \url{https://github.com/JordanAwan/DP_KStests}.

\subsection{Goodness-of-fit tests}\label{app:good}
The KS and Kuiper tests are straightforward to implement. Besides numerically evaluating the integral, we were unaware of a more efficient method of implementing the Wasserstein test. We include in this section a simple algorithm for Cram\'er-von Mises.

\subsubsection{Cram\'er-von Mises test}
While the Cram\'er-von Mises test is phrased in terms of an integral, when applied to goodness-of-fit tests, there is a simple algorithm to calculate the test statistic. Because the test statistic is continuous-valued, we add Laplace noise rather than Tulap noise. 
\begin{algorithm}[H] 
\caption{Private Cram\'er-von Mises}
\label{alg:Cramer}
\scriptsize
INPUT: $x\in \RR^n$, ${F}$, and $\epsilon>0$
\begin{algorithmic}[1]
  \setlength\itemsep{0em}
  \STATE $\omega^2 = \frac{1}{12n} + \sum_{i=1}^{n}(\frac{2i-1}{2n}-{F}(x_i))^2$ where $x_1$,...$x_n$ are sorted in increasing order 
  \STATE $\twid T =  \sqrt{\frac{\omega^2}{n}} +\frac{1}{n}N$, where $N\sim \mathrm{Laplace}(1/\epsilon)$
\end{algorithmic}
OUTPUT: $\twid T$
\end{algorithm}

\subsection{Two sample tests}\label{app:two}

\subsubsection{Kruskal Wallis Test}
\citet{couch2019} modified the Kruskal-Wallis test statistics by measuring the absolute value of distance between two samples. They proved that the sensitivity of the modified test is bounded by 8 and showed that it has higher power than the rank-based Kruskal-Wallis test.
\begin{algorithm}[H] 
\caption{Private Kruskal-Wallis Test \cite{couch2019}}
\scriptsize
INPUT: $x\in \RR^{n}$, $y\in \RR^m$, and $\epsilon>0$
\begin{algorithmic}[1]
  \setlength\itemsep{0em}
  \STATE Set $N=n+m$
  \STATE $n_1 = \mathrm{length}(x)$, $n_2 = \mathrm{length}(y)$
  \IF{n is even}
    \STATE $H = \frac{4(N-1)}{N^2}\left(n\left|\sum_{i=1}^n \mathrm{rank}(x_i) -\frac{n+1}{2}\right|+m\left|\sum_{j=1}^m \mathrm{rank}(y_j)-\frac{m+1}{2}\right|\right)$, where $\mathrm{rank}$ computes the rank in the combined sample $(x,y)$.
  \ELSIF{n is odd}
    \STATE H =  $\frac{4}{N+1}\left(n\left|\sum_{i=1}^n \mathrm{rank}(x_i)-\frac{n_{1}+1}{2}\right|+m\left|\sum_{j=1}^m \mathrm{rank}(y_j)-\frac{n_2+1}{2}\right|\right)$
  \ENDIF 
  \STATE $ \tilde{H} = H + 8N,$, where $N\sim \mathrm{Laplace}(1/\epsilon)$
\end{algorithmic}
OUTPUT: $\tilde{H}$
\end{algorithm}

\subsubsection{Private Mann-Whitney test}
We modified the DP Mann Whitney test statistics proposed from \citet{couch2019}, 
because the group size is assumed to be known in our settings. Our modification results in less noise added to make for a fairer comparison with the other tests. 
\begin{algorithm}[H] 
\caption{Private Mann-Whitney Test \citep{couch2019}}
\scriptsize
INPUT: $x\in \RR^n$, $y\in \RR^m$, and $\epsilon>0$
\begin{algorithmic}[1]
  \setlength\itemsep{0em}
   \STATE $U_1 = \sum_{i=1}^n \mathrm{rank}(x_i), U_2 = \sum_{i=1}^m \mathrm{rank}(y_j)$, where $\mathrm{rank}$ computes the rank in the combined sample $(x,y)$.
  \STATE $U = \min\{U_1, U_2\}$
  \STATE $ \twid U = U + \max\{m,n\}N$, where $N\sim\mathrm{Laplace}(1/\epsilon)$
\end{algorithmic}
OUTPUT: $\twid U$
\end{algorithm}

\subsubsection{Private median test}
\citet{awan2018differentially} gives a DP median test to test whether the paired samples have the same median. 
\begin{algorithm}[H] 
\caption{Private Median Test \citep{awan2018differentially}}
\scriptsize
INPUT: $x, y\in \RR^n$, and $\epsilon>0$
\begin{algorithmic}[1]
  \setlength\itemsep{0em}
 \STATE $S = \#\{i \mid \mathrm{rank}(x_i)>n\}$, where $\mathrm{rank}(x_i)$ is the rank in the combined sample $[x,y]$.
  \STATE $\twid S = S + N$, where $N\sim \mathrm{Tulap}( \exp(-\epsilon),0)$
  \STATE $\twid T = |\twid S-\frac{n}{2}|$
\end{algorithmic}
OUTPUT: $\twid T$
\end{algorithm}

\subsection{Paired data tests}\label{app:paired}
\subsubsection{Sign Test}
The sign test is designed to determine whether two paired samples have the same median. \citet{awan2018differentially} defined the DP version of sign test by adding a Tulap random variable. The algorithm assumes that there are no ties $x_i= y_i$.

\begin{algorithm}[H] 
\caption{Private sign test \citep{awan2018differentially}}
\scriptsize
INPUT: $x,y\in \RR^n$, and $\epsilon>0$
\begin{algorithmic}[1]
  \setlength\itemsep{0em}
  \STATE $T = \# \{x_i > y_i\}$
  \STATE $\twid T = T + N$, where $N \sim \mathrm{Tulap}(\exp(-\epsilon),0)$
\end{algorithmic}
OUTPUT: $\twid T$
\end{algorithm}

\subsubsection{Wilcoxon signed-rank test}
Given the datasets $(x_i$, $y_i)$, the Wilcoxon signed-rank test evaluates whether the difference between two independent datasets is symmetric around 0. The version of Wilcoxon test used by \cite{couch2019} was introduced in \citet{pratt1959}: Given two paired datasets $x_i$, $y_i$, the algorithm computes the difference $d_i$ between them. Compared with original Wilcoxon test, it keeps the rows with $d_i$ = 0 and set $s_i$ = 0, and then rank them by their magnitude. 

\begin{algorithm}[H]  
\caption{Private Wilcoxon Test \citep{couch2019}}
\scriptsize
INPUT: $x,y\in \RR^n$, and $\epsilon>0$
\begin{algorithmic}[1]
\STATE $d_i = y_i-x_i$
\STATE $W = \sum_{i=1}^n \mathrm{rank}(|d_i|)*\mathrm{sign}(d_i)$ 
\STATE $\twid W  = W + (2n)N$, where $N\sim \mathrm{Laplace}(1/\epsilon)$
\end{algorithmic}
OUTPUT: $\twid W$
\end{algorithm}
  

\bibliographystyle{imsart-nameyear} 
\bibliography{ref.bib}       


\end{document}